\documentclass[twocolumn,twocolappendix]{aastex631}
\usepackage{aas_macros}
\usepackage{float}
\usepackage{amsmath,amssymb}
\usepackage{natbib} % For use with bibtex
\usepackage{graphicx} % For included graphics
\usepackage{color}
\usepackage{verbatim}
\usepackage{rotating}

\newcommand{\proptosim}{\mathrel{\vcenter{
 \offinterlineskip\halign{\hfil$##$\cr
 \propto\cr\noalign{\kern2pt}\sim\cr\noalign{\kern-2pt}}}}}
\renewcommand{\b}[1]{\boldsymbol{#1}}
\renewcommand{\i}{\ensuremath{\rm i}}
\newcommand{\unit}[1]{{\rm\, #1}}

\newcommand{\response}[1]{{#1}}

\renewcommand{\min}{\mathrm{min}}
\renewcommand{\max}{\mathrm{max}} 
\newcommand{\au}{\mathrm{AU}}
\newcommand{\cm}{\unit{cm}}
\newcommand{\m}{\unit{m}}
\newcommand{\g}{\unit{g}}
\newcommand{\G}{\unit{G}}
\newcommand{\K}{\unit{K}} 
\newcommand{\km}{\unit{km}}
\newcommand{\kms}{\unit{km\ s^{-1}}}

\renewcommand{\micron}{\unit{\mu m}}

\newcommand{\erg}{\unit{erg}}
\newcommand{\eV}{\unit{eV}}
\newcommand{\keV}{\unit{keV}}
\newcommand{\s}{\mathrm{s}}
\newcommand{\yr}{\mathrm{yr}}

\newcommand{\ang}{\ensuremath{\mathrm{\AA}}}

\newcommand{\lya}{{\text{Ly}\ensuremath{\alpha}}}

\newcommand{\B}{\mathbf{B}}     % Magnetic fields
\newcommand{\E}{\mathbf{E}}     % Electric fields
\newcommand{\J}{\mathbf{J}}     % Electric current fields
\newcommand{\F}{\mathbf{F}}     % Fluxes
\renewcommand{\v}{\mathbf{v}}   % Velocity fields
\renewcommand{\d}{\mathrm{d}}
\newcommand{\e}{\mathrm{e}}

\renewcommand{\ion}[2]{{\rm#1}\;\textsc{#2}}
\newcommand{\p}{\mathrm{p}}     % Poloidal
\newcommand{\eq}{\mathrm{eq}}   % equal-temperature
\renewcommand{\O}{\mathrm{O}}     % Ohmic
\newcommand{\A}{\mathrm{A}}     % Ambipolar
\renewcommand{\H}{\mathrm{H}}     % Hall
\renewcommand{\P}{\mathrm{P}}     % P
\newcommand*\chem[1]{\ensuremath{\mathrm{#1}}}
\renewcommand{\i}{\ensuremath{\mathrm{i}}}
\usepackage[colorinlistoftodos]{todonotes}

\renewcommand{\B}{{\rm B}}

\newcommand{\figdir}{.}

\begin{document}

\title{Photoevaporation from Inner Protoplanetary Disks
  Confronted with Observations}

\author[0000-0003-2509-6558]{Yiren Lin}
\affil{The Kavli Institute for Astronomy and Astrophysics,
  Peking University, Beijing 100871, China}
\affil{Department of Astronomy, School of Physics, Peking
  University, Beijing 100871, China}
  
\author[0000-0002-6540-7042]{Lile Wang}
\affil{The Kavli Institute for Astronomy and Astrophysics,
  Peking University, Beijing 100871, China}
\affil{Department of Astronomy, School of Physics, Peking
  University, Beijing 100871, China}
  
\author[0000-0001-8060-1321]{Min Fang}
\affil{Purple Mountain Observatory,
  Chinese Academy of Sciences, 10 Yuanhua Road, Nanjing
  210023, China}
\affil{University of Science and Technology
  of China, Hefei 230026, China}

\author[0000-0002-9220-0039]{Ahmad Nemer}
\affil{Center for Astrophysics and Space Science, 
New York University Abu Dhabi}

\author[0000-0002-6710-7748]{Jeremy Goodman}
\affil{Department of Astrophysical Sciences, Princeton
  University, Princeton NJ 08540, USA}

\correspondingauthor{Lile Wang} 
\email{lilew@pku.edu.cn}

\begin{abstract}
  The decades-long explorations on the dispersal of
  protoplanetary disks involve many debates about
  photoevaporation versus magnetized wind launching
  mechanisms. This work argues that the observed
  winds originating from the inner disk
  ($R\lesssim 0.3~\au$) cannot be explained by the
  photoevaporative mechanism. Heating the gas to proper
  temperatures for the observed forbidden lines (especially
  [\ion{O}{i}] $6300~\ang$) will over-ionize it, suppressing
  the abundances of species responsible for the
  emission. Even if adequate emissivity is achieved by
  fine-tuning the physical parameters, the total cooling
  power will become unattainable by the radiative heating
  alone. Energy conservation requires the presumed
  photoevaporative winds to be heated to $\gtrsim 10^5~\K$
  when launched from inner disks. However, due to efficient
  thermal accommodation with dust grains and cooling
  processes at high densities, X-ray irradiation at energies
  above $1~\keV$ cannot efficiently launch winds in the
  first place because of its high penetration. Some studies
  claiming X-ray wind launching have oversimplified the
  thermochemical couplings. Confirmed by semi-analytic
  integrations of thermochemical fluid structures, such high
  ionizations contradict the observed emission of neutral
  and singly-ionized atoms from the winds originating from
  the inner disks.
\end{abstract}


\section{Introduction}
\label{sec:intro}

As the birthplaces of planets, protoplanetary disks (PPDs)
undergo dispersal through various processes, including
planet formation, accretion onto the central protostar, and
outflowing in winds. The latter two processes compete with
planet formation, limiting the available time and mass for
it \citep{2023ASPC..534..567P}. Photoevaporation occurs in
the absence of magnetic fields, where high-energy photons
heat the gas and unbind it from the star
\citep[e.g.][]{2006MNRAS.369..216A, 2009ApJ...690.1539G,
  2010MNRAS.401.1415O, 2017ApJ...847...11W}. These winds
only carry off their own share of angular momenta and are
not directly related to disk accretion. On the other hand,
magnetized winds can exert torque on the disk and drive
accretion. Two types of magnetized winds have been
identified: magnetocentrifugal winds and magneto-thermal
winds; modeling suggests that the latter type prevails in
PPDs \citep{2016ApJ...818..152B}. Numerical studies of
magnetized winds and associated accretion demand
the consistent inclusion of physical processes related to
non-ideal magnetohydrodynamics (MHD) \citep[e.g.][]
{Wardle+Konigl1993, Bai+Goodman2009, 2016ApJ...819...68X,
  2017ApJ...845...75B, 2019ApJ...874...90W}. With proper
treatments of radiation and subsequent non-equilibrium
thermochemistry, ``hybrid'' winds with
  photoevaporation based on and spatially located above
magnetized outflows were found in
\citet{2019ApJ...874...90W}, yet the foundations were still
the magnetized winds.

Forbidden line emissions from atoms and ions (e.g.,
[\ion{O}{i}] $6300~\ang$, and [\ion{Ne}{ii}] $12.8~\micron$)
have been used as prospective indicators of PPD
winds. Researchers have proposed expected observables and
synthetic observations accordingly, based on the
photoevaporative wind models \citep[e.g.][]
{2006MNRAS.369..229A, 2008MNRAS.391L..64A,
  2016MNRAS.460.3472E}, as well as the magnetized wind
models \citep{2020ApJ...904L..27N}. Emerging observations
with high spatial and spectral resolution indicate that the
broad and generally (and usually slightly)
blueshifted low-velocity component (LVC) of [\ion{O}{i}]
$6300~\ang$ emission come from regions with relatively high
densities of neutral materials \citep{2019ApJ...870...76B,
  2023ApJ...945..112F} and fairly close to the central star
($R\lesssim 0.3~\au$ for TW Hya, a solar-mass
star, \citealt{2023NatAs...7..905F}). These results are in
favor of the magnetized wind models as they are sufficient
to provide adequate physical conditions without being
limited by the depths of gravitational potential wells.

% However, arguments from the photoevaporation side also
% emerge, claiming that photoevaporative winds are also able
% to explain what has been observed \citep[e.g.][R23
% hereafter] {2023ApJ...955L..11R}.  This letter focuses on
% the discussion about why photoevaporation is insufficient
% to launch outflows and produce the [\ion{O}{i}] emission
% lines from the inner disk, thus emphasizing the necessity
% of the magnetized mechanisms in the study of PPD
% evolution.

Arguments from the photoevaporation side also emerge. Some
models, with an emitting and mostly static or circulating
corona above the inner disk and photoevaporation winds
launched at larger radii, were proposed to claim that
photoevaporative winds are also able to explain what has
been observed \citep[e.g.][R23 hereafter]
{2023ApJ...955L..11R}. However, a circulating corona of gas
that does not drive materials to infinity is not
  an outflow, and should not be defined as a
  ``photoevaporative wind''. Discussions about wind
  launching mechanisms should not rely on the analyses of
  inner-disk coronae that are not winds.  Meanwhile, the
  line profiles of the [\ion{O}{i}] 6300 emission of various
  PPDs are generally wide ($\gtrsim 20~\kms$) and often
  blueshifted \citep{2018ApJ...868...28F}. Such line
  profiles are the indications of outflows, rather than the
  hydrostatic or circulating gas above disk surfaces. This
problem is especially unresolvable if the basic assumption
about photoevaporation, that the dynamics are driven almost
solely by stellar radiation, is made for an inner-disk
corona, no matter if it is hydrostatic, in circulation
motion, or inside a wind.  In what follows, we will use the
word ``corona'' to denote the coronal region above the disk
surface in general, no matter it is a photoevaporative wind,
hydrostatic gas, or the coronal gas in
circulating motion.

This paper focuses on demonstrating the conflicts
between the photoevaporative disk wind model and
observation results, based on the detailed observations on
TW Hya \citep{2023NatAs...7..905F} in particular, while
also applicable to a broader range of PPDs as well: the
inner disk within $\sim 1~\au$ from the host star could
not launch photoevaporation with physical consistency,
given the observation constraints especially on the
luminosity of [\ion{O}{i}] $6300~\ang$ emission. Similar
concerns have been raised since the discoveries of PPD
outflows \citep[e.g.][] {Font+2004}, and this work
discusses the general conditions of PPD wind launching
from the inner disk using latest observational constrains.
The paper is organized as follows. The required
thermal conditions concerning neutral oxygen radiation,
cooling power, and photoevaporation are described
in Section \ref{sec:chemistry}, which is one of the core
issues conflicting photoevaporation with
observation. Section \ref{sec:energetics} discusses the
issue regarding the energetic balance of the inner-disk
photoevaporative wind model. Section \ref{sec:ana} presents
more comprehensive semi-analytic models that exposes those
issues quantitatively. The overall conclusions are
summarized and discussed in Section \ref{sec:summary}.
Details of several physics issues relevant to the
discusses are elaborated in the Appendices.

% \response{ The following discussions will show that
%   fine-tuning does not help, since the photoevaporation
%   scenario is physically untenable.  We will discuss about
%   why photoevaporation is insufficient to produce the
%   compact [\ion{O}{i}] emission (\S\ref{sec:chemistry}) or
%   even launch winds from the inner disk
%   (\S\ref{sec:energetics}).  The arguments are supplemented
%   with results from a semi-analytic model in
%   \S\ref{sec:ana}.  In \S\ref{sec:summary}, we discuss the
%   problems with photoevaporation models and emphasize the
%   necessity of the magnetized mechanisms in the study of PPD
%   evolution.  }

\section{Ionization and Emissivity}
\label{sec:chemistry}

The emission of the [\ion{O}{i}] $6300~\ang$ line requires
neutral oxygen atoms existing at relatively high
temperatures ($\gtrsim 5000~\K$, but not too hot--otherwise
oxygen will be ionized thermally). How much neutral oxygen
is needed?  For TW Hya, \citet{2023NatAs...7..905F} reported
the luminosity on the emission line
$L_{6300} \sim 1.5 \times 10^{- 5} L_{\odot} \sim 5.8 \times
10^{28}~\erg~\s^{-1}$.
% which further gives the emissivity of this
% line,
% \begin{equation}
%   \varepsilon \simeq \frac{h c}{\lambda}
%   n_{\chem{O}} x_u A \sim 2 \times 10^{-12}
%   ~\erg~\cm^{-3}~\s^{-1}  x_u \left(
%     \frac{n_{\chem{O}}}{10^2~\cm^{-3}} \right) .
% \end{equation}
% When the number density of electron is much smaller than
% the critical density
% $n_{\rm {crit}} \equiv A / k_{\rm de}$ and the temperature
% is not too high ($T_4 \lesssim 2$) so that most oxygen
% atoms are neutral, the equation above can be solved to
% obtain,
% \begin{equation}
%   \begin{split}
%     & x_u \sim \frac{g_l k_{\rm {ex}} n_e}{A}, \\
%     & \varepsilon  \simeq \frac{h c}{\lambda}
%       n_{\chem{O}} x_u A 
%       \simeq 10^{-12}~\erg~\cm^{-3} ~\s^{-1} \\
%     & \quad \times \left(
%     \frac{n_{\chem{O}}}{10^2~\cm^{-3}} \right) \left(
%     \frac{n_e}{10^6~\cm^{-3}} \right) \exp
%       \left( - \frac{2.2838}{T_4} \right) .
%   \end{split}
% \end{equation}
The volume of a sphere with $R = 0.5~\au$ is
$1.8 \times 10^{39}~\cm^3$, which sets the lower limits of
the emissivity,
$\varepsilon\gtrsim 3 \times 10^{-11}~\erg~\cm^{-3}
~\s^{-1}$.  Considering the typical value of elemental
abundance of oxygen as
$X_{\chem{O}} \sim 3 \times 10^{- 4}$, such a high
emissivity will require oxygen to be predominantly neutral
inside the wind if the electrons with density
$n_e \gtrsim 10^6~\cm^{-3}$ all come from the ionization of
hydrogen, and $T \gtrsim 10^4~\K$.

\begin{figure}
 \centering 
 \includegraphics[width=8.6cm]{\figdir/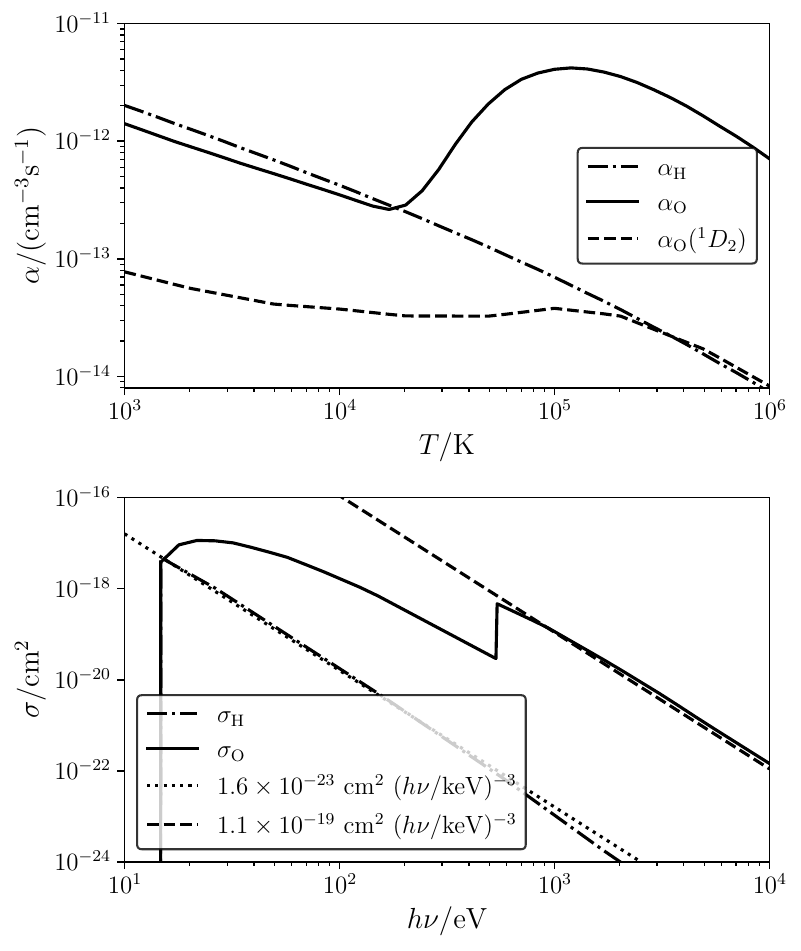}
 \caption{ Key atomic data adopted in this work.  {\bf Upper
     panel}: Recombination rates as functions of
   temperatures (based on \citealt{2004A&A...425.1153M,
     adas}), presenting the total case B recombination rates
   $\alpha_{\rm H}$ for \ion{H}{ii} (dash-dotted curve),
   $\alpha_{\rm O}$ for \ion{O}{ii} (solid curve), and
   $\alpha_{\rm O}(^1D_2)$ specifically for the
   recombination to ${}^1D_2$ state of \ion{O}{i} (dashed
   curve; populations of \ion{O}{ii} parent states have been
   considered). These rates have both radiative and
   dielectronic recombination components summarized.  {\bf
     Lower panel}: Photoionization cross sections of
   \ion{H}{i} (dashed curve) and \ion{O}{i} (solid curve)
   based on \citet{Verner+etal1996}. The power-law
   approximations are also presented for hydrogen (dotted
   curve) and oxygen ($h\nu > 530~\eV$; dashed curve). }
 \label{fig:alpha-oxygen}
\end{figure}

\subsection{Ionization conditions}
\label{sec:chem-ionization}

\begin{figure*}
    \centering,
    \includegraphics[width=18cm]{\figdir/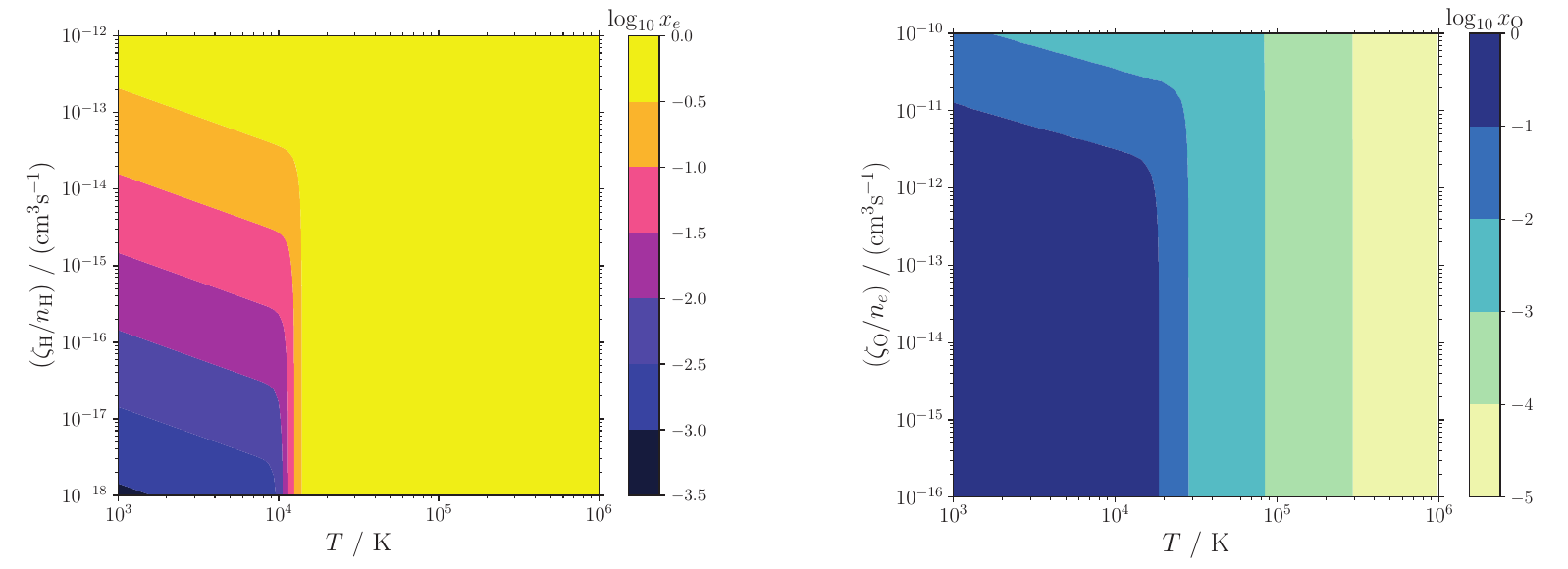}
    \caption{Contour plots for the ionized fraction of
      hydrogen (left) and neutral fraction of oxygen
      (right), as functions of $(T/\K)$ and
      $(\zeta_{\chem{H}} / n_{\chem{H}}) /
      \left(\cm^3~\s^{-1} \right)$ (left panel) or
      $(\zeta_{\chem{O}} / n_e) / \left(\cm^3~\s^{-1}
      \right)$ (right panel). 
      For $T \gtrsim 3 \times 10^4~\K$, the ionized fraction is close to 1 and the neutral fraction of oxygen drops to $x_{\chem{O}} < 10^{-2}$.
      %Note that the parameter spaces
      %that yield relatively high $x_e$ and sufficient
      %$x_{\rm O}$ almost do not overlap (see also
      %\S\ref{sec:chem-thermal-ion}).
      }
   \label{fig:chem-ionization}
\end{figure*}

\subsubsection{Lengths and timescales of photoionization}

One of the most important destroyers of neutral oxygen is
high-energy photons.  The ``standard'' X-ray luminosity used
in R23 is
$L_X = 2 \times 10^{30}~\erg~\s^{-1}\simeq 5\times
10^{-4}~L_\odot$ (also adopted in
\citealt{2017ApJ...847...11W, 2019ApJ...874...90W}).
% which leads to photoionization rates of oxygen quantified
% in eq.~\eqref{eq:ene-ionize}.
The coronal region above the disk surface within
$R = 1~\au$ standing between the source and wind bases at
larger distances, either a hydrodynamic wind or a
hydrostatic corona, must be transparent to high-energy
photons, otherwise those photons will not reach the disk
surface to launch the wind.

The X-ray ionization cross sections of atomic hydrogen and
oxygen are, above the inner shell ionization threshold
($h\nu\gtrsim0.53~\keV$; see also the lower panel of
Figure~\ref{fig:alpha-oxygen}), 
\begin{equation}
  \label{eq:ene-sigma-ph-ion}
  \begin{split}
     & \sigma_{\chem{H}, i} \sim 1.6 \times 10^{-23}~\cm^2
      \times \left( \frac{h \nu}{\keV} \right)^{-3} ,\\
    & \sigma_{\chem{O}, i} \sim
      1.1\times10^{-19}~\cm^2 
      \times
      \left( \frac{h \nu}{\keV} \right)^{-3}\ ,
  \end{split}
\end{equation}
which can be applied to estimate the photoionization rate of
oxygen and hydrogen,
$\zeta_{Z}\simeq L_X\sigma_{Z,i}/(4\pi R^2h\nu)$ ($Z$ is
either \chem{O} or \chem{H}).  Constrained by the
observations, the wind velocity \footnote{\response{The wind
    velocity observed in TW Hya \citep{2023NatAs...7..905F}
    is $v_{\rm w} \sim1~\km~\s^{-1}$ for [\ion{O}{i}] and
    $v_{\rm w} \sim 5~\km~\s^{-1}$ for [\ion{Ne}{ii}]. We
    adopt the latter, which is more likely representative of
    inner disk winds in general.}}  is at the order of
$v_{\rm w} \sim 5~\km~\s^{-1}$,
%(as adopted by R23)
leading to the following estimate of the attenuation length
of neutral oxygen abundances,
\begin{equation}
  \label{eq:chem-len-oxy-neutral}
  \begin{split}
    & \lambda_{\chem{O}, i} \sim \zeta_{\chem{O}}^{-1} v_{\rm w}
      \sim 0.062~\au \times \left( \frac{L_X}{2\times
      10^{30}~\erg~\s^{-1}} \right)^{-1} \\
    &\quad \times
      \left(\frac{v_{\rm w}}{5~\km~\s^{-1}} \right)
      \left(\frac{R}{0.3~\au} \right)^{2}
      \left( \frac{h \nu}{\keV}\right)^{4} .
  \end{split}
\end{equation}
In other words, for X-rays with photon energy $1~\keV$, the
abundance of neutral oxygen will decrease by one decade
every $(\ln\,10)\times \lambda_{\chem{O}, i} \sim 0.14~\au$
as the gas traverses the wind. Only a few times $0.1~\au$
above the wind base, the neutral fraction of oxygen becomes
too low. 
% Note also that, even if an important fraction of neutral
% oxygen and hydrogen atoms manage to survive in the inner
% disk corona, their absorption of ionizing photons will
% inhibit XUV from irradiating and evaporating the outer
% disk. This shielding effect undermines the physical
% picture proposed in R23, which has a static inner corona
% with a fraction of neutral oxygen and a photoevaporating
% outer disk.

\subsubsection{Ionization and recombination}
\label{sec:chem-thermal-ion}

In view of the short
% \response{attenuation length for ionizing photons}
ionizing length for the outflowing gas,
ionization equilibrium should establish quickly; hence a
lower bound to the ionization fraction can be obtained by
balancing \emph{collisional} ionization against
recombination. The recombination rate coefficient for the
reaction $\ion{O}{ii}+ e^- \rightarrow \ion{O}{i}$
is described by the total rates in
Figure~\ref{fig:alpha-oxygen}. The corresponding
collisional ionization rate
$\ion{O}{i} + e^- \rightarrow 2 e^- + \ion{O}{ii}$
\citep{Bell+etal1983}, we have (subscript ``ci''
for collisional ionization; $T_4 \equiv T / 10^4~\K$),
% $I_{\chem{O}} / k_B = 158053~\K$,
\begin{equation}
  \label{eq:chem-ion-oxy}
  % \begin{split}
  %   \alpha_{\chem{O}}
  %   & \simeq  3.2 \times 10^{-13}~\cm^3~\s^{-1} 
  %     T_4^{-0.66},\\
    k_{\rm O, ci}
     \simeq 8.7 \times 10^{-9}~\cm^3~\s^{-1} T_4^{1/2}
      \exp \left( -\dfrac{15.81}{T_4} \right).
  % \end{split}
\end{equation}
The ionization condition of oxygen is bound by the electron
abundance, constrained by the ionization balance of
hydrogen, where the recombination rate is also presented in
Figure~\ref{fig:alpha-oxygen}, and the 
%recombination 
collisional ionization rate
approximately reads,
\begin{equation}
  \label{eq:chem-ion-hyd}
  % \begin{split}
  %   \alpha_{\chem{H}}
  %   & \simeq 2.5 \times 10^{-13}~\cm^3~\s^{-1} T_4^{-0.75},\\
  k_{\chem{H}, {\rm ci}}
  \simeq 6.4 \times 10^{-9}~\cm^3~\s^{-1}  T_4^{1/2}
  \exp \left( -\dfrac{15.78}{T_4} \right) .
      %     \end{split}
\end{equation}
Although charge exchange processes,
$\chem{O}^+ + \chem{H} \leftrightarrow \chem{O} +
\chem{H}^+$, may be important in non-equilibrium
calculations, their effects are readily covered by detail
balances in equilibrium states. Even in non-equilibrium
calculations, at sufficiently high temperatures
($T \gg 228~\K$, which corresponds to the difference in
ionization energy between $\chem{H}$ and $\chem{O}$), the
charge exchange has negligible impacts on the ionization
conditions.

Let $n_{\chem{H}}$ be the total number density of hydrogen $n_{\chem{H}} = n (\chem{H}) + n (\chem{H}^+)$, $x_e$ be the ionized fraction of hydrogen $x_e \equiv n_e / n_{\chem{H}}$, and $n (\chem{H}) = (1 - x_e) n_{\chem{H}}$. We first solve the ionization balance of hydrogen
%Using $n_{\chem{H}} = n (\chem{H}) + n (\chem{H}^+)$, we
%first solve the ionization balance of hydrogen, using
%$x_e \equiv n_e / n_{\chem{H}}$ and
%$n (\chem{H}) = (1 - x_e) n_{\chem{H}}$ 
(the negative branch of solution is ignored):
\begin{equation}
  \label{eq:chem-bal-elec}    
  % \begin{split}
  % &
  \alpha_{\chem{H}} n_{\chem{H}}^2 x_e^2 -
      \zeta_{\chem{H}} n_{\chem{H}} (1 - x_e) -
     k_{\rm H, ci} n_{\chem{H}}^2 x_e (1-x_e)= 0\ .
    % \\
    % & x_e = [2 (\alpha_{\chem{H}} + k_{\rm H, ci})]^{-1}
    %   \times \bigg\{ k_{\chem{H}, {\rm ci}} -
    %   \left(\dfrac{\zeta_{\chem{H}}}{n_{\chem{H}}}\right) \\
    % &\ \ + \left[k_{\chem{H}, {\rm ci}}^2 + 2 
    %   \left(\dfrac{\zeta_{\chem{H}}}{n_{\chem{H}}}\right) 
    %   (2 \alpha_{\chem{H}} + k_{\rm H, ci}) + 
    %   \left(\dfrac{\zeta_{\chem{H}}}{n_{\chem{H}}}\right)^2
    %   \right]^{1/2}\bigg\} .
  % \end{split}
\end{equation}
% Taking eq.~\eqref{eq:ene-ionize} into the solution
% (eq.~\ref{eq:chem-bal-elec}), one obtains the ionized
% fraction of hydrogen shown in the left panel of
% Figure~\ref{fig:chem-ionization}.
Given the electron number density $n_e$, the ionization
balance of oxygen reads (where $x_{\chem{O}}$ is the
fraction of neutral oxygen compared to total oxygen, $x_{\chem{O}}\equiv n(\chem{O})/n_{\chem{O}}$),
\begin{equation}
  \label{eq:chem-bal-oxy}    
  % \begin{split}
  % &
  \alpha_{\chem{O}} n_e n_{\chem{O}} (1 - x_{\chem{O}}) -
    \zeta_{\chem{O}} n_{\chem{O}} x_{\chem{O}} -
    k_{\chem{O}, {\rm ci}} n_e n_{\chem{O}}x_{\chem{O}}= 0\ .
  %   \\
  % & x_{\chem{O}} = \dfrac{\alpha_{\chem{O}} n_e}
  %   {(\alpha_{\chem{O}} + k_{\chem{O}, {\rm ci}}) n_e +
  %   \zeta_{\chem{O}} } .
  % \end{split}
\end{equation}

% Assuming $\zeta_{\chem{O}} \rightarrow 0$ to stay on the
% safe side (which is equivalent to $n_e \rightarrow \infty$),
% the fraction reduces to,
% \begin{equation}
%   \label{eq:chem-bal-oxy}
%   \begin{split}
%     x_{\chem{O}}
%     & \simeq \dfrac{1}{1 + k_{\chem{O}, {\rm
%       ci}}/\alpha_{\chem{O}}} \\
%     & \simeq \left[ 1 + 2.7\times 10^4\ T_4^{1.16} 
%   \exp \left( -\frac{15.8}{T_4} \right) \right]^{-1}.
%   \end{split}
% \end{equation}
% which is in accordance with the Saha equation.

Solutions to eqs.~\eqref{eq:chem-bal-elec},
\eqref{eq:chem-bal-oxy} under different conditions are
presented in Figure~\ref{fig:chem-ionization}.  When
$T \gtrsim 3 \times 10^4~\K$, the neutral fraction of oxygen
drops to $x_{\chem{O}} < 10^{-2}$. The neutral fraction
will drop even further down to $x_{\chem{O}} < 10^{-4}$
% with $n_e \sim 10^2~\cm^{-3}$
at $T \sim 10^5~\K$. Such low abundances of
neutral oxygen is generally inadequate to yield the observed
[\ion{O}{i}] $6300~\ang$ emission luminosity.  As
we will also discuss later in
\S\ref{sec:ene-potential}, the specific thermal energy
corresponding to this temperature is also insufficient to
drive outflows even from $R = 1~\au$, let alone
$R = 0.3~\au$ or $0.1~\au$.

From Figure~\ref{fig:chem-ionization}, one can infer that
the biggest hope of obtaining significant fractions of
neutral oxygen is to stand close to the lower-left corner,
i.e. with low temperatures ($T$) and weak
ionization-electron ratio ($\zeta_{\chem{O}} /
n_e$). However, attempts in this domain will inevitably bog
down to the following multilemma. One way to lower
$\zeta_{\chem{O}} / n_e$ is to have small
$\zeta_{\chem{O}}$, which means either small $L_{\rm X}$ or
strong attenuation of X-ray. Both possibilities will inhibit
photoevaporative winds, as this wind launching mechanism
requires intensive radiative heating. Another way to lower
$\zeta_{\chem{O}} / n_e$ is to have greater $n_e$. Then here
comes another problem: where do the electrons come from?
\begin{itemize}
  \item If they come mainly from the photoionization of
    hydrogen (including secondary ionization by electrons or
    recombination photons in the X-ray ionization cascades),
    then the photon flux should be large enough to ionize
    oxygen to a greater extent--remember
    $\sigma_{\chem{O}, i}$ is significantly greater than
    $\sigma_{\chem{H}, i}$ for X-ray photons ($\sim 10^4$
    times at $h \nu = 1~\keV$), due to the interactions with
    inner shell electrons.
  \item If they are produced via collisional ionization,
    then the temperature should be high enough to
    significantly ionize hydrogen--this will also lead to
    significant ionization of oxygen, as their ionization
    energy values are almost the same.
\end{itemize}
These issues narrow the possible parameter space for the
co-existence of \ion{O}{i}, sufficient free
electrons, and relatively high temperatures; all of which
are necessary for sufficiently luminous [\ion{O}{i}]
$6300~\ang$ radiation.

\subsection{Are sufficient emissivities possible?}
\label{sec:chem-higher-emis}

Given the Einstein $A$ coefficient
$A \simeq 5.63 \times 10^{-3}~\s^{-1}$ \citep{NIST_ASD} of
this transition from $^1 D_2$ to $^3 P_2$, the collisional
excitation (subscript ``ex'') and de-excitation (``de'')
rates are approximately (see also
\citealt{1985ApJ...291..722T, Pequignot1990}; subscripts
``\chem{H}'' and ``$e$'' indicate the collision partner,
neutral hydrogen or free electron),
%$T_4 \equiv T / 10^4~\K$
\begin{equation}
  \label{eq:chem-coll-rates}
  \begin{split}
    & k_{{\rm de}, e} \simeq 4.1 \times
      10^{-9}~\cm^3~\s^{-1} \times 
      \left( \frac{T_4^{0.93}}{1 + 0.605 T_4^{1.105}} \right),
    \\
    & k_{\rm de, H} \simeq 5.7 \times
      10^{-13}~\cm^3~\s^{-1},
    \\
    & k_{{\rm ex, H}, e} \simeq k_{{\rm de, H}, e}
      \frac{g_u}{g_l} \exp
      \left( - \frac{h c / \lambda}{k_\B T} \right)  .
  \end{split}
\end{equation}
The population fraction of the upper $^1 D_2$
state $x_u$ and the lower $^3 P_2$ state $x_l$ can be
estimated by, assuming $x_u + x_l \simeq 1$ and
$n(\H) = n_\H (1 - x_e)$,
\begin{equation}
  \label{eq:chem-upper-state}
  \begin{split}
    &\quad n(\chem{O}) x_u [ k_{{\rm de}, e} n_e +
      k_{\rm de, H} n(\H) + A] \\
    & \simeq n(\chem{O}) x_l
      [ k_{{\rm ex}, e} n_e + k_{\rm ex, H} n(\H) ]
      + \alpha_{\chem{O}}(^1D_2) n(\chem{O^+}) n_e ;
      \\
    & x_u \simeq
    \dfrac{k_{{\rm ex},e} + k_{\rm ex, H}(x_e^{-1}-1)
      + \alpha_{\chem{O}}(^1D_2) (x_\chem{O}^{-1}-1) } 
      { (k_{{\rm ex},e}+k_{{\rm de},e}) + (x_e^{-1}-1)
      (k_{\rm ex, H}+k_{\rm de, H}) + A n_e^{-1} } ,
  \end{split}
\end{equation}
where the $\alpha_{\chem{O}}(^1D_2)$ values adopted are
presented in the upper panel of
Figure~\ref{fig:alpha-oxygen}, based on the atomic physics
data in \citet{2004A&A...425.1153M, adas}. Throughout the
temperature range concerned, one can observe that
$\alpha_{\chem{O}}(^1D_2) < 10^{-13}~\cm^3~\s^{-1}$ even
with the dielectronic recombination and the cascade from
higher recombination products taken into account. The
collisional excitation by neutral hydrogen $k_{\rm ex, H}$
has considerable contribution to the emissivity only when
$T< 10^4~\K$.  The solution ignoring these two terms,
$x_u\simeq k_{{\rm ex}, e} / (k_{{\rm ex},e} + k_{{\rm
    de},e} + A n_e^{-1})$, is a reasonable approximation
throughout most of the parameter space concerned.

\begin{figure*}
  \centering,
  \includegraphics[width=18cm]
  {\figdir/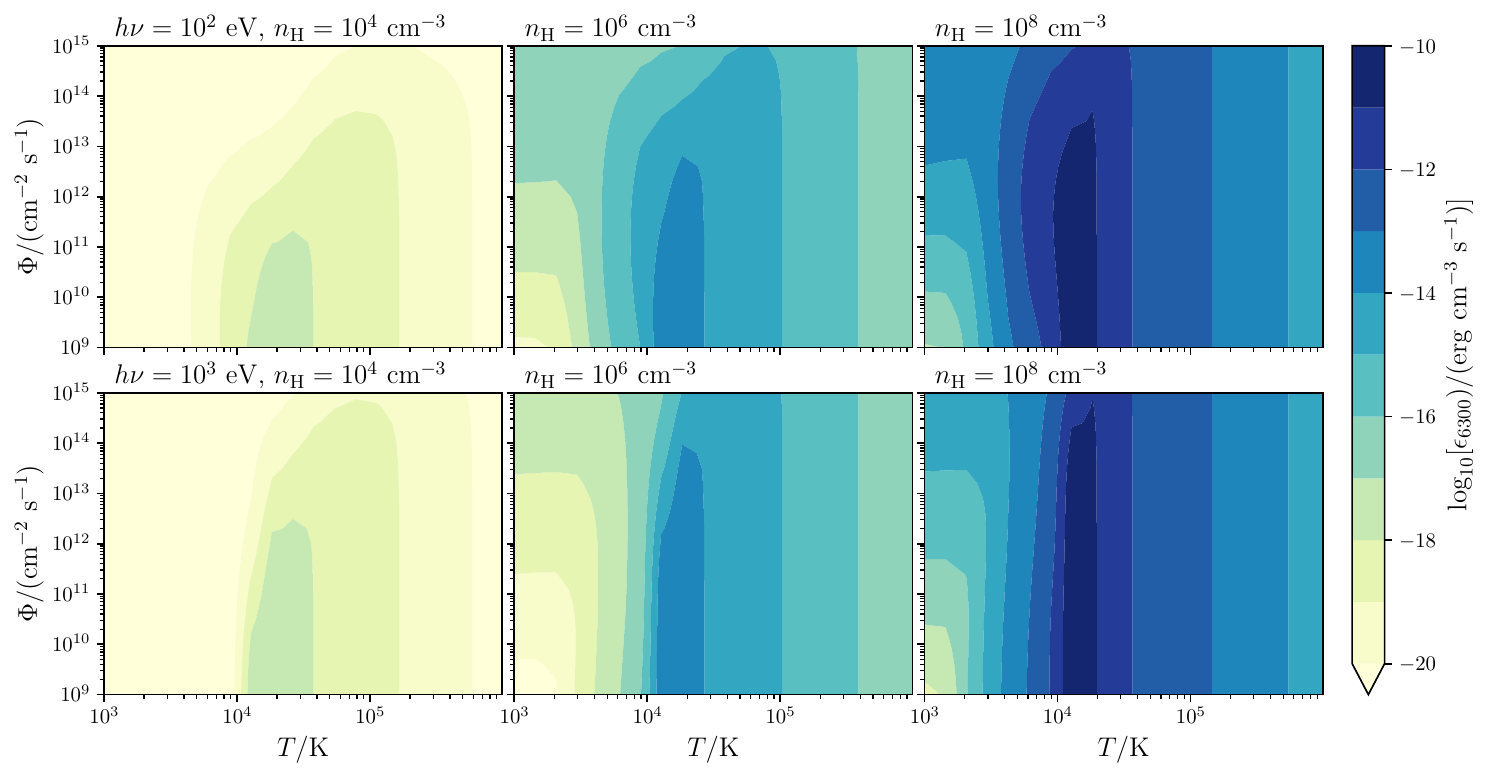}
  \caption{Emissivity of the [\ion{O}{i}] $6300~\ang$ line
    under different radiation fluxes ($\Phi$ for the fluxes
    of ionizing photons) and temperatures, calculated by
    solving eqs.~\eqref{eq:chem-upper-state}
    (\S\ref{sec:chem-higher-emis}). The upper row presents
    the results for ionizing photon energy
    $h\nu = 10^2~\eV$, and the lower row for
    $h\nu = 10^3~\eV$. Different gas density values,
    quantified by the total hydrogen nuclei number density
    $n_\H$, are used in different columns as denoted. }
  \label{fig:chem-emis}
\end{figure*}

Using the population number given by
eqs.~\eqref{eq:chem-upper-state}, the emissivity through the
[\ion{O}{i}] line can be estimated by
$\epsilon_{6300} = n_{\rm H} X _{\rm O} x_{\rm O} x_u A
h\nu_{6300}$ ($h\nu_{6300} = 1.97~\eV$ is the photon energy
of the [\ion{O}{i}] $6300~\ang$ emission; not to be confused
with the ionizing photon energy $h\nu$). The results are
presented in Figure~\ref{fig:chem-emis}, showing highly
insufficient emissivity for the
$n_{\rm H} \leq 10^6~\cm^{-3}$ cases. The
$n_{\rm H} \gtrsim 10^8~\cm^{-3}$ cases obtain regions in
the parameter spaces that yield sufficient
$\epsilon_{6300}$, at temperatures
$1\lesssim T_4 \lesssim 2$.  These parameters are quite
common in magnetized disk wind models with reasonable
selection of physical parameters
\citep[e.g.][]{2015ApJ...801...84G, 2017ApJ...845...75B,
  2019ApJ...874...90W, 2020A&A...639A..95R}. In contrast,
gas heated solely by radiation, either in photoevaporative
winds or hydrostatic coronae, will have difficulties to
maintain these physical conditions.

Gas at high temperature ($T_4> 1$) and high density
($n_\H> 10^8~\cm^{-3}$) simultaneously is susceptible to
significant cooling. The total cooling power in the emitting
region reads approximately,
\begin{equation}
  \label{eq:chem-emis-cooling}
  \begin{split}
    & L_{\rm cool} \simeq V_{6300}\Lambda n_e n_{\rm H} 
      \ \simeq 5\times 10^{-3} L_\odot \times x_e
      \left(\dfrac{V_{6300}}{0.5~\au^3}\right) 
      \\
    & \times\left(\dfrac{\Lambda}
      {10^{-24}~\erg~\cm^3~\s^{-1}}\right)
      \left(\dfrac{n_{\rm H}}{10^{8}~\cm^{-3}}\right)^2.
      % \left(\dfrac{x_e}{10^{-1}}\right) .
  \end{split}
\end{equation}
Here $V_{6300}$ is the volume of the emitting region (the
volume of an $r = 0.5~\au$ sphere is $\sim 0.5~\au^3$), and
$\Lambda$ is the cooling rate coefficient. When the gas has
relatively high ionization fraction ($x_e\gtrsim 10^{-1}$),
the cooling via case B recombination of hydrogen alone
yields $\Lambda > 3\times
10^{-25}~\erg~\cm^{-3}~\s^{-1}$. When photoionization is not
overwhelming, Lyman-$\alpha$ cooling gives
$\Lambda> 10^{-23}~\erg~\cm^3~\s^{-1}$ if optically thin
\citep[e.g.][]{2012ApJS..202...13G}, or effectively
$\Lambda\gtrsim 10^{-25}~\erg~\cm^3~\s^{-1}$ if optically
thick (see also the Monte Carlo simulations for optically
thick Lyman-$\alpha$ radiative transfer with similar gas
densities, e.g. \citealt{2017MNRAS.472.2773G}). Inclusion of
other cooling mechanisms will only make $L_{\rm cool}$
greater. Such a dense and hot corona will radiate at power
$L_{\rm cool} \gtrsim 10^{-3}~L_\odot$, which requires fairly
powerful heating comparable to the total X-ray power of
typical T Tauri stars. For magnetized disk winds that can
power the wind gas the extraction of disk gravitational
energy via accretion, the upper limit of power injected into
the winds, $P_\max$, approximately reads,
\begin{equation}
  \label{eq:chem-mhd-power-max}
  \begin{split}
    P_\max
    & \sim \dfrac{G M_* \dot{M}_{\rm acc}}{R}
    \\
    & \sim 3\times 10^{-3}~L_\odot
      \left( \dfrac{\dot{M}_{\rm acc}}
      {10^{-8}~M_\odot~\yr^{-1}} \right)
      \left( \dfrac{R}{0.5~\au} \right)^{-1}\ ,
  \end{split}
\end{equation}
where $M_*$ is the mass of the protostar,
$\dot{M}_{\rm acc}$ is the accretion rate, and the $R$
here describes the typical radial location of wind gas
heating. While the magnetized wind is sufficient to power a
hot and dense corona, radiation alone seems questionable.

Is it possible to maintain the temperature of such a dense
corona by stellar radiation alone? Optical and infrared
radiation, carrying most energy from the central protostar,
cannot heat the gas beyond the blackbody temperature of the
stellar photosphere due to the second law of
thermodynamics. What is more, if an $r \sim 0.5~\au$,
$T\sim 10^4~\K$ gas sphere is optically thick in optical and
infrared bands, one can get, by the Stefan-Boltzmann law, a
ridiculous $L_{\rm cool}\sim 10^5~L_\odot$ cooling
luminosity. High-energy photons face a different dilemma. If
the radial optical depth is low for a band of radiation,
energy deposition for photons in this band will
be very inefficient. On the other end, high optical depths
prevent photons from reaching anywhere beyond the innermost
part of the corona. This will stop the radiative heating for
most of the coronal gas, and removes the energy source for
photoevaporative winds at larger disk radii described in
R23. The radial column density of an $n_\H = 10^8~\cm^{-3}$
coronal gas over an $0.5~\au$ radial line is
$N_\H \sim 10^{21}~\cm^{-2}$, which is too high for EUV if
the neutral fraction is $(1-x_e)\gtrsim 10^{-3}$, and too
low otherwise (for EUV with very low neutral fraction, and
for X-ray and FUV). The only possibility is to restrict the
physical parameters so that the radial optical depth of the
corona is at the order of unity for X-ray, yet this will
raise the cooling power to $L_{\rm cool}\gtrsim L_\odot$ due
to the increase of $n_\H$. More generally speaking, such
fine-tuning significantly undermines the capability and
robustness of any theoretical models.

\section{Energetics of the Outflows}
\label{sec:energetics}

In addition to the thermochemical issues, the
energy balance also challenges the mechanisms of wind
launching from the inner disk. As multiple
observations have figured out, the low-velocity components
(LVC) of [\ion{O}{i}] $6300~\ang$ emission from PPDs are
often blueshifted \citep[e.g.][]{2018ApJ...868...28F,
  2019ApJ...870...76B, 2023ApJ...945..112F,
  2023NatAs...7..905F}. This is a strong indication that the
emission regions are located inside outflows rather than a
hydrostatic corona.
% Energetics of outflows should also be examined to
% determine the driving mechanisms of the winds.  It is
% worth noting that, in case that the emitting gas cannot
% reach \response{infinity} (in other words, the streamlines
% threading through the emitting gas fall back to the disk),
% the dynamics of gas should be described as
% ``circulations'' instead of ``outflows'', and the wind
% launching mechanism cannot be concluded as
% photoevaporation.

In general, gas has to be heated sufficiently to escape
from the stellar potential well.
% Although some early works declared that photoevaporative
% winds can still be launched with $c_s^2 < G M_* / 2 R$
% ($M_*$ is the stellar mass, $R$ is the cylindrical radius
% coordinate), any detailed inspections will indicate that
% this is unphysical.
As a reference, the wind mass-loss rate $\dot{M}$
is approximately related to the wind parameters (here $\mu$
is the mean molecular mass, $v_{\rm w}$ is the wind speed,
and $n_\chem{H}$ is the total number density of
hydrogen nuclei),
\begin{equation}
  \label{eq:ene-mass-loss}
  \begin{split}
    \dot{M}
    & \sim 0.4 \times 10^{-9}~M_{\odot}~\yr^{-1}\\
    & \times \left(
      \frac{n_\chem{H}}{10^7~\cm^{-3}}\right) 
    \left( \frac{\mu}{m_p} \right)
    \left( \frac{R}{\au} \right)^2
    \left( \frac{v_{\rm w}}{5~\km~\s^{-1}} \right) .
  \end{split}
\end{equation}

% \response{This section examines the energetics of the
% outflow to determine if photoevaporation is sufficient to
% drive the winds from the inner disk.  We estimate the
% heating due to X-ray photons(\S\ref{sec:ene-heating}) and
% compare it with the gravitational
% potential(\S\ref{sec:ene-potential}). In
% \S\ref{sec:ene-problems} we discuss related issues with
% photoevaporation models such as R23.}

\subsection{Gravitational potential and the escape}
\label{sec:ene-potential}

Is the heating enough for gas to escape the stellar
potential well? We begin with the assumption of
instantaneous injection of radiation energy for simplicity
and a more realistic scenario will be elaborated in
\S\ref{sec:ana}.

Under the hydrodynamic conditions (Appendix~\ref{sec:mfp}),
the velocity of fluid particles is thermalized, and the
thermal energy has to be greater than the potential well
depth to escape. Considering the fact that the gas in an
protoplanetary disk orbits the central star at a velocity
very close to the local Keplerian speed
$v_\K\equiv (G M_*/R)^{1/2}$, the effective depth
of the potential well satisfies ($\varepsilon_k$ is the
specific kinetic energy),
\begin{equation}
  \label{eq:ene-specific-pot}
  \begin{split}
    |\phi|
    & \simeq | \Phi + \varepsilon_k |
      \simeq \left| - \frac{G M_*}{R} + \frac{v_\K^2}{2}
      \right| = \frac{G M_*}{2 R} \\
    & \simeq 4.6~\eV~m_p^{-1} \times
      \left( \frac{R}{\au} \right)^{-1}
      \left( \frac{M_*}{M_{\odot}} \right) .
  \end{split}
\end{equation}
Even if significant flaring raises the
altitude of the wind base, one can easily prove that,
\begin{equation}
  \label{eq:ene-pot-flared}
| \phi |_{\theta} = \left| - \frac{G
M_*}{R / \cos \theta} + \frac{v_\K^2}{2} \cos^3 \theta
\right| = | \phi | \cos \theta (1 + \sin^2 \theta) ,
\end{equation}
where $\theta$ is the flaring angle (the angle between the
vector from the star to the wind base and the equatorial
plane). Compared to more complete calculations
that involve consistent azimuthal velocity $v_\phi$, this
estimation is still approximately valid. The relative
deviation from the mid-plane Keplerian speed is roughly
$\Delta v_\phi / v_\K \simeq -(h/R)^2(z/h)^2/2 $
\citep[e.g.][]{2002ApJ...581.1344T}, whose absolute value
is less than $10~\%$ near the flared disk surface where
$z/r\sim 4$. In addition, the consistent $v_\phi$ is
slower than $v_\K$, and the estimation in
eq.~\eqref{eq:ene-pot-flared} stands on the safe side by
  underestimating $| \phi |_{\theta}$. The factor
$f_\theta\equiv \cos \theta (1 + \sin^2 \theta)$ satisfies
$f_\theta \in(7/8,\ 4\sqrt{6}/9)$ when $\theta\in[0,\pi/3]$,
and decreases to $f_\theta < 0.5$ only when $\theta >
1.3$. Disk flaring does {\it not} make the potential well
appreciably shallower, because the steady-state specific
kinetic energy also decreases at higher altitudes. Given the
depth of the potential well, the analysis of energy
constraints on the heated outflows can be qualitatively
divided into two limiting cases: ballistic motion (or most
equivalently, adiabatic), and isothermal.

% \subsubsection{Ballistic motion}

On the ballistic motion limit, an adiabatic parcel of heated
gas receives no more energy injection after
heating processes near the wind bases. When it tries to
escape from the surface of a protoplanetary disk at
$R = 1~\au$, the specific thermal energy of the gas must be
at least $\varepsilon_g > | \phi | = 4.6~\eV /
m_p$. Considering fully ionized hydrogen plasma, this is
equivalent to $\gtrsim 2.3 ~ \eV$ energy per particle due to
equipartition of energy, which is equivalent to
$3 k_\B T / 2 \gtrsim 2.3 ~ \eV$, or
$T \gtrsim 1.8 \times 10^4~\K$. This temperature
significantly suppresses the neutral fraction of oxygen down
to $\sim 10^{-2}$ even in absence of photoionization (see
discussions in \S\ref{sec:chemistry}), leading to the
insufficiency of neutral oxygen atoms.  When coming to the
inner sub-$\au$ region as indicated by
\citet{2023NatAs...7..905F}, high temperatures at least
$T \gtrsim 1.8 \times 10^5~\K$ are required. At such
temperatures, $x_{\chem{O}}$ will be suppressed down to
$x_{\chem{O}} \lesssim 3 \times 10^{- 4}$ (see also
Figure~\ref{fig:chem-ionization}). Note also that the
discussions are about fully ionized plasma with small
$\mu$. In case of neutral or even molecular gas discussed in
R23, the situation will become even worse.

One may be concerned by the photons in the X-ray band, which
seem to have greater photon energies than the potential well
depths. As we elaborate in Appendix~\ref{sec:ene-heating},
photons are even more incapable of heating the gas to the
temperature required by the photoevaporative outflows due to
their high penetration.  \response{ One might also be
  curious about why outflows may still be driven (although
  with tiny mass-loss rates) from the inner disk in
  photoevaporative conditions still seem to be launched in
  some photoevaporation models, given the potential well
  depths.}
  % (e.g., \citealt{2016MNRAS.460.3472E}, R23)
We discuss this issue in Appendix~\ref{sec:ene-problems},
emphasizing that the oversimplified treatments of gas
thermodynamics (by mapping the temperature to the ionization
parameter) could always lead to wind launching even when the
heated gas is not sufficiently energetic to escape.

\subsection{Isotropic isothermal outflows}
\label{sec:ene-isotropic}

The other limit is an isothermal outflow, which
often results from efficient heating and thermal transfer.
While the escape of adiabatic gas requires
$|\phi| \lesssim k_BT/\mu$, isothermal coronae always
launch outflows \citep[e.g.][]{1958ApJ...128..664P}. The
key to estimate the isothermal wind mass-loss rate is the
mass density at the sonic critical point $\rho_s$ using
the isothermal sound speed $c_s$ and the mass density at
the wind base $\rho_b$, 
\begin{equation}
  \label{eq:ene-isotherm-rhos}
  \dfrac{\rho_s}{\rho_b}
  \sim\exp\left(-\frac{|\phi_b|}{c_s^2}\right)
  \simeq\exp\left(-\frac{GM_*}{r_b c_s^2}\right)\ ,
\end{equation}
which can be derived from the hydrostatics of
gravitated isothermal spheres. This is a reasonable
approximation in the subsonic regions, and the error of
estimated mass-loss rates is no more than $\sim 50~\%$ as
one can verify with the standard Parker wind mode. The
sonic critical radius roughly reads
$r_s\sim GM_*/(2c_s^2)\simeq 5~\au$ for gas at $10^4~\K$
escaping from a solar-mass star. The mass-loss rate is
related to $\rho_s$ by,
\begin{equation}
  \rho_s\simeq\frac{\dot{M}}{4\pi r_s^2 c_s}\simeq
  10^{-18}~\g~\cm^{-3}\times
  \left(\dfrac{\dot{M}}{10^{-9}~M_\odot~\yr^{-1}}\right) .
\end{equation}
% or equivalently, $n_{\chem{H},s}\simeq10^6~\cm^{-3}$.
Using eq.~\eqref{eq:ene-isotherm-rhos}, one gets
$n_{\chem{H},b}=10^{10\sim11}~\cm^{-3}$ for $r_b\sim 1~\au$,
or $n_{\chem{H},b}\sim 10^{21}~\cm^{-3}$ for $r_b\sim
0.3~\au$. While the former is on the high end of the
possible range, the latter is far beyond any realistic
PPD gas.

\section{Comprehensive Semi-analytic Models}
\label{sec:ana}

% Analyses in \S\ref{sec:energetics} may have one caveat:
% energy injection takes place locally, and may not
% accomodate for the changes in circumstances when gas
% moves outwards. This section combine the mechanical and
% thermochemical considerations with a non-local
% scheme. Admittedly the full set of physical processes
% involved in the photoionized gas overwhelms the capacity
% of analytic solutions. We decide to stand on the safe side
% when simplifications are mandatory.

% The analyses in \S\ref{sec:energetics} focus on the
% energetics of the whole wind, without taking much care
% about the motion of the fluids itself. This approach may
% have one limitation: the injection of energy occurs
% locally and may not account for the changes in
% circumstances when gas move outward.  This section
% integrates mechanical and thermochemical equations with a
% global approach, guided by the typical disk
% photoevaporation geometry portrayed in
% Figure~\ref{fig:ana-schematic}. Formulations of the
% equations are elaborated in Appendix~\ref{sec:ana-appdx}.
% It is important to note that the complete set of physical
% processes involved in the photoionized gas surpasses the
% capabilities of analytic solutions. Therefore, in case
% that anything has to be neglected, we choose to err on the
% safe side.

The analyses in \S\ref{sec:chemistry} and
\S\ref{sec:energetics} have discussed the energy issues from
both aspects of thermochemistry and hydrodynamics,
respectively. These approaches may have a limitation: the
localized energy injection and subsequent ballistic motion
of gas may not fully account for the changes in
circumstances that occur when gas moves outwards.  This
section integrates the physics of both therochemistry and
hydrodynamics for more comprehensive discussions, guided by
the typical geometry of disk photoevaporation depicted in
Figure~\ref{fig:ana-schematic}.

For simplicity, we adopt the estimation method
for hydrodynamic outflow profiles based on the hydrostatic
equations. This approach has been adopted by various early
works on stellar outflows, % (inspired by e.g.,
% \citealt{1958ApJ...128..664P})
and here we show that such estimations err on the safe
side. Consider the radial momentum equation along a ``flux
tube'' spanned by a streamline and its neighbors (see
Figure~\ref{fig:ana-schematic}), under the steady state
[$\partial(\rho v)/\partial t\equiv0$],
\begin{equation}
  \label{eq:radial-mom}
  \partial_R p = -\rho v_R \partial_R v_R -
  \dfrac{GM_*\rho}{R^2}\ .
\end{equation}
With $v_R > 0$ and $\partial_R v_R > 0$ as the gas
accelerates before reaching the sonic point, the
hydrodynamic solutions always have more negative
$\partial_R p$ than the hydrostatic solutions with
$v_R = 0$. By integrating from the same wind launching
point to the same radius $R$ for $p$, the pressure profile
estimated by hydrostatics is always higher than the actual
hydrodynamic values. If one assumes that the outflows are
isothermal, the hydrostatic approximation (through
multiplying the density at sonic point by the sound speed)
will overestimate the mass-loss rate
($\dot{M}_{\rm stat} > \dot{M}_{\rm dyn, iso}$). Once the
themodynamics inside outflows are closer to adiabatic
rather than isothermal, the actual mass-loss rate will be
even lower than the isothermal case
($\dot{M}_{\rm dyn, iso} > \dot{M}_{\rm dyn}$). In other
words, if the hydrostatic estimations yield a small
mass-loss rate, one should expect that the actual wind
mass-loss rate is even lower. 

\subsection{Formulations of the Coronal hydrostatics}
\label{sec:ana-eqs}

\begin{figure}
 \centering 
 \includegraphics[width=6.5cm]{\figdir/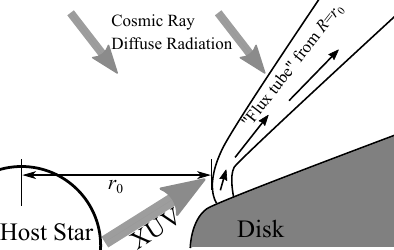}
 \caption{Schematic setups for the semi-analytic models
   (\S\ref{sec:ana}), showing an XUV irradiated ``flux
   tube'' originating from $R=r_0$.  }
 \label{fig:ana-schematic}
\end{figure}

% In absence of outflows, hydrostatic equilibrium models
% properly describe the structures of the heated disk
% corona.  Even with outflows, fluids are still in
% hydrostatic equilibrium below the wind base.
Figure~\ref{fig:ana-schematic} illustrates that the
fluid geometry can be described by fluid ``flux tubes'',
which have approximately the same solid angle at different
radii, for both the static fluids stalled above a point of
the disk surface, and the wind launched from a point on the
disk. The guiding equations can be written for the
hydrostatic with thermochemistry and radiation in one
dimension in spherical geometry,
\begin{equation}
  \label{eq:hydrostat-dimensional}
  \begin{split}
    & 0 = \partial_R p + \frac{G M_* m_\H n}{R^2}\ ,
    \\
    & p = n (1 + x_e) k_\B T\ ,
    \\
    & 0 = \alpha_\H x_e^2 - \frac{F\sigma_{\H,i}}{n}
      (1 - x_e) - k_{\rm H,ci} x_e (1 - x_e)\ ,
    \\
    & 0 = \frac{F\sigma_{\H,i}}{n} (1 - x_e) (h \nu -
      I_\H) - \Lambda\ ,\\
    & 0 = \partial_R\ln F + \left[ \frac{2}{R} + n
     \sigma_{\H,i} (1 - x_e) \right] .    
  \end{split}
\end{equation}
Here, $F$ is the photon number flux of the XUV
photons. Eqs.~\eqref{eq:hydrostat-dimensional} are solved in
their dimensionless forms with the following dimensionless
transforms (where we use
$n_0 \equiv 1 / (\sigma_{\H,i} r_0)$ for the reference
number density, and $k_0 \equiv 10^{-8}~\cm^3~\s^{-1}$ for
the fiducial rate coefficient),
\begin{equation}
  \begin{split}
    & \zeta \equiv \frac{R}{r_0},\
      \varrho \equiv \frac{n}{n_0},\
      \varpi \equiv \frac{p r_0}{G M_* m_\H n_0},\
      \Theta \equiv \frac{k_\B T}{I_\H},\\
    & \varphi \equiv \frac{F\sigma_{\H,i}}{n_0 k_0},\
      \epsilon\equiv \frac{h \nu}{I_\H},\ 
      \Gamma \equiv \frac{I_\H r_0}{G M_* m_\H} .
  \end{split}
\end{equation}
The collisional ionization and recombination rates are
reduced to dimensionless functions (see also
eqs.~\ref{eq:chem-ion-hyd}),
\begin{equation}
  \begin{split}
    & k_{\rm H,ci} \equiv k_0 \kappa (\Theta)\ ,\
      \kappa(\Theta) \simeq 2.3\ \Theta^{1 / 2} \e^{- 1 /
      \Theta}\ ;
    \\ 
    & \alpha_\H \equiv k_0 A (\Theta)\ ,\ A (\Theta)
      \simeq 3.2 \times 10^{- 6} \ \Theta^{-0.75}\ .
      % Not the most accurate form but to be consistent with
      % \label{eq:chem-ion-hyd}
  \end{split}
\end{equation}
The cooling term is decomposed into three components:
collisional ionization cooling ($\Lambda_{\rm ci}$),
recombination cooling ($\Lambda_{\rm rec}$), and Lyman
$\alpha$ cooling ($\Lambda_\lya$). These terms are also
reduced into their dimensionless forms (see also
\citealt{2009ApJ...693...23M}, \citealt{Black1981}),
\begin{equation}
  \label{eq:hydrostat-cooling}
  \begin{split}
    \Lambda_{\rm rec}
    & \equiv \lambda_{\rm rec} (\Theta) x_e^2 \times
      I_\H k_0\ ,
    \\
    \Lambda_\lya
    & = \lambda_\lya(\Theta) x_e(1-x_e) \times I_\H k_0\ ,
    \\ 
    \Lambda_{\rm ci}
    & = \kappa (\Theta) x_e (1 - x_e) \times I_\H k_0\ ;
    \\
    \lambda_\lya (\Theta)
    & \simeq \frac{3.44}{1 + 1.26\Theta^{1 / 2}} \exp \left(
      - \frac{3}{4 \Theta} \right)\ ,
    \\ 
    \lambda_{\rm rec} (\Theta)
    & \simeq (0.569 - 0.0416 \ln \Theta)\;\Theta\;
      A (\Theta) \ .
  \end{split}
\end{equation}
The total cooling is decomposed into
$\Lambda = \Lambda_{\rm ci} + \Lambda_{\rm rec} + \beta
\Lambda_\lya$, where the $\beta$ factor is the escape
probability of cooling \lya\ photons. It is also noted that
the set of cooling processes involved is far from being
complete. By setting a default $\beta\rightarrow 0$ and
ignoring all other cooling processes, the gas becomes less
bound, making them easier to escape and putting us on the
safe side by exaggerating the effect oppositing to our
conclusions.

With such transforms, the dimensional
eqs.~\eqref{eq:hydrostat-dimensional} are recast into the
dimensionless form,
\begin{equation}
  \label{eq:hydrostat-dimensionless}
  \begin{split}
    \partial_{\zeta} \varphi
    & = - [2 \zeta^{- 1} + \varrho (1-x_e)] \varphi\ ,
    \\
    \partial_{\zeta} \varpi
    & = - \zeta^{-2} \varrho\ ,
    \\
    0 & = \varphi \varrho^{-1} (1-x_e) + \kappa (\Theta) x_e
        (1 - x_e) - A (\Theta) x_e^2\ ,
    \\
    0 & = \varphi \varrho^{- 1} (1-x_e) (\epsilon - 1) -
        \lambda_{\rm rec} (\Theta) x_e^2
    \\
    &\qquad -[\kappa(\Theta)+\beta\lambda_\lya(\Theta)]
      x_e (1-x_e)\ , 
    \\
    \varpi & = \Gamma (1+x_e) \varrho \Theta\ .
  \end{split}
\end{equation}
Two extra constraints are required when solving the algebraic
part of these equations.  First, the carbon element should
be predominantly ionized, due to the FUV photons (with
$h\nu>11.2~\eV$) from the central star and the diffuse
interstellar radiation fields. A lower bound
$x_{e,\min} = X_\chem{C}$ should be imposed, and we choose
$X_\chem{C}=1.4\times 10^{-4}$. Second, the optical and
infrared radiation from the host star also imposes a lower
limit on the gas temperature,
\begin{equation}
  \label{eq:min-T}
  T_\min \simeq T_{\rm eq} = 886~\K\times
  \left(\dfrac{L_*}{L_\odot}\right)^{1/4}
  \left(\dfrac{R}{0.1~\au}\right)^{-1/2} .
\end{equation}
Here $T_{\rm eq}$ is the equilibrium temperature of black
bodies. We choose the bolometric luminosity $L_*=L_\odot$ in
this work, and determine the lower bound for $T$ (and for
$\Theta$ subsequently) according to the distance to the
star. Meanwhile, the ionization by cosmic ray can also be
non-negligible when the density is sufficiently low, and we
include an extra $\zeta_{\rm CR} = 10^{-17}~\s^{-1}$ as
another ionization source in addition to the central star
irradiation.

\subsection{Properties of the Solutions}
\label{sec:ana-sol-prop}

Eqs.~\eqref{eq:hydrostat-dimensionless} % are a set of
% (ordinary) differential-algebraic equations, and
can be solved with semi-analytic methods by selecting a set
of physical parameters and inner boundary conditions at
$r_0$. The solutions can be roughly categorized into three
types: % according to their outflow properties:
\begin{itemize}
\item Gravitationally bound. When the injected irradiation
  has tiny thermochemical impacts, the situation is similar
  to an externally gravitated polytrope: all gas is
  gravitationally bound and has no outflows, the system will
  have a surface (at which the gas density and pressure
  vanish) at a finite radius.
  % and can be adequately described by
  % eqs.~\eqref{eq:hydrostat-dimensionless}.
\item Pressure bound. When the gas pressure converges to a
  constant value (``terminal pressure'') at infinite
  distances (similar to isothermal atmospheres gravitated by
  the central star), and such a constant value is lower than
  the ISM pressure
  ($\sim 4\times 10^{-13}~{\rm dyn~cm}^{-2}$ for both warm
  and cool neutral media; see
    \citealt{Jenkins+Tripp2011}).
  % Based on the typical interstellar media properties
  % ($T\sim 5000~\K$ and $n\sim 0.6~\cm^{-3}$ for the warm
  % neutral media, or $T\sim 100~\K$, $n\sim 30~\cm^{-3}$
  % for the cool neutral media; see also
  % \citealt{DraineBook}), we pick the dimensional critical
  % pressure
  % $p_\crit \sim 4\times 10^{-13}~{\rm dyn~cm}^{-2}$. When
  % an analytic model has dimensional pressure $p < p_\crit$
  % at infinity, it will also be treated as a bound model
  % the system is also contained by its sourroundings.
\item Unbound. When the terminal pressure is greater than the
  surroundings.%, outflows are inevitable.
\end{itemize}

When the gas is bound, the gas cannot reach
infinity and eventually falls back to the disk, which is
not a ``wind'' or ``outflow'' by definition. In the
unbound case, one should in principle connect a steady-state
hydrodynamic solution to a hydrostatic solution at a
specific point (often called ``wind base''). Nevertheless,
such problems are always underdefinite when connecting a
hydrostatic solution to hydrodynamic solutions analytically
\citep[e.g.][] {2018ApJ...860..175W}, and the hydrodynamic
part is always numerically stiff and less
trustworthy. Therefore, we estimate the most important
parameter for outflows--the mass-loss rates--by
$\dot{M} = 4\pi r_s^2c_s\rho_s$, in which $r_s$, $c_s$, and
$\rho_s$ are the radius, sound speed, and mass density at
the sonic critical radius where the radial outflow becomes
supersonic. Since the actual mass density at the critical
radius is certainly less than the hydrostatic estimation
because of gas acceleration, this simplification is
over-estimating the mass-loss. Such $\dot{M}$ is an
extrapolation of a flux tube to the full $4\pi$ solid angle,
which also lead to an over-estimation due to the geometry
(see also Figure~\ref{fig:ana-schematic}).

For a perfectly isothermal Parker wind
\citep{1958ApJ...128..664P}, $r_s$ is located at
$r_s = GM_*/2c_s^2$. The models described by
eqs.~\eqref{eq:hydrostat-dimensionless} have variable
temperatures. We thus define a dimensionless Bernoulli
parameter,
\begin{equation}
\label{eq:bernouli}
\mathcal{B} \equiv \gamma\varpi/(\gamma - 1) -
\varrho/\zeta
\end{equation}
here $\gamma$ is the adiabatic index and
$\gamma = 5/3$ in all relevant cases in this work. When a
solution is gravitationally bound, its $\mathcal{B}$ is
always negative. The unbound solutions, in contrast, have
transition points from negative $\mathcal{B}$ to positive
$\mathcal{B}$, and we approximate the sonic critical point
and associated physical quantities by this transition. Note
that there are also cases whose $\mathcal{B}$ becomes
positive but pressure converges to very low values at large
radii.

\subsection{Ionized Corona Solutions}
\label{sec:ana-sol}

\begin{figure}
 \centering 
 \includegraphics[width=8cm]{\figdir/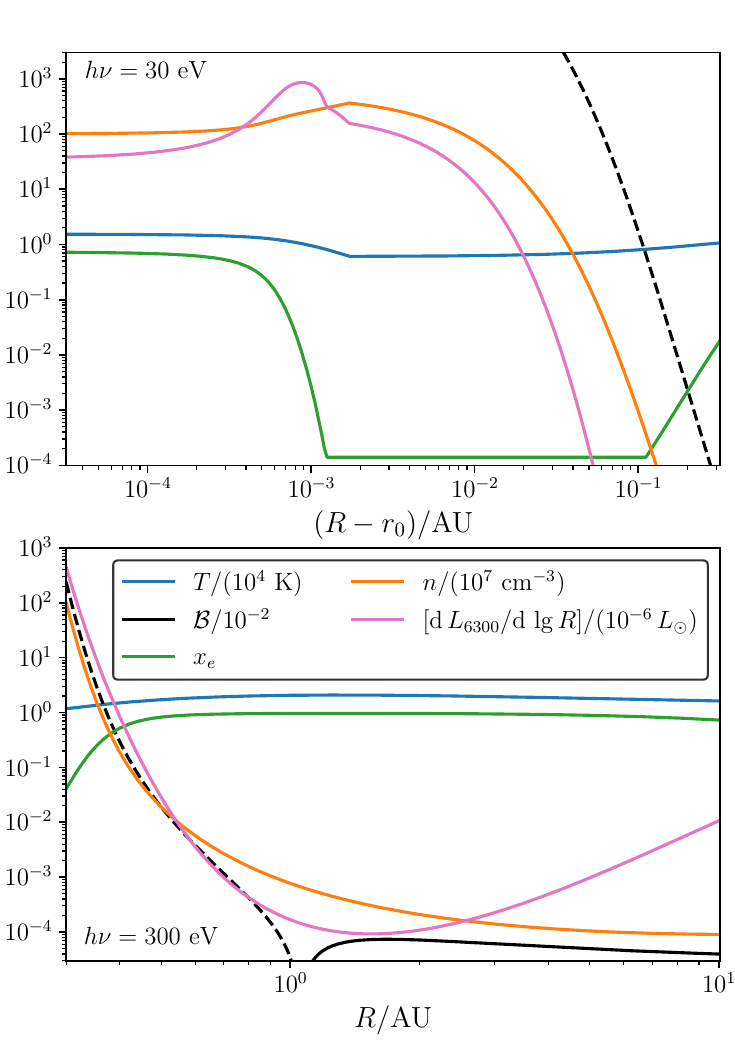}
 \caption{Example solutions for a bound model (upper
   panel; note that its horizontal axis is $R-r_0$) and an
   unbound model (lower panel), showing the hydrostatic
   and thermochemical profiles (distinguished by
   colors). The [\ion{O}{i}] emissions are presented in the
   luminosity per logarithm radius. Both models have the
   same key parameters ($M_* = M_\odot$, $r_0 = 0.3~\au$,
   $n_0=10^9~\cm^{-3}$, $L_{\rm XUV} = 10^{-3}\;L_\odot$),
   with one exception: the bound model has XUV photon
   energy $h\nu = 30~\eV$, while the unbound model has
   $h\nu = 300~\eV$. The Bernoulli parameters
   (\S\ref{sec:ana-sol-prop}) are presented in dashed
   lines when negative and in solid lines when positive.}
 \label{fig:ana-profiles}
\end{figure}

Two solutions with typical selections of physical parameters
are presented in Figure~\ref{fig:ana-profiles} showing an
unbound solution with XUV (X-ray and extreme ultraviolet)
photon energy $h\nu = 300~\eV$, and a bound solution with
$h\nu = 30~\eV$. The unbound model exhibits a sign
transition in the Bernoulli parameter $\mathcal{B}$ at
$R\sim 1~\au$ (eq.~\ref{eq:bernouli}), at which the mass
density already drops by a factor $\lesssim 10^{-5}$
compared to $n_0$. The sound speed at that location,
corresponding to the temperature $T\sim 2\times 10^4~\K$, is
around $\sim 12~\km~\s^{-1}$, yet the low mass density leads
to a negligible mass-loss rate
$\dot{M}\lesssim 10^{-12}~M_\odot~\yr^{-1}$. To
quantify the significance of wind mass loss, the local
wind mass-loss timescale $\tau_{\rm wind}$ is defined as,
\begin{equation}
  \tau_{\rm wind} \sim
  \dfrac{\d M/\d \ln R}{\d \dot{M}/\d \ln R} 
  \simeq \dfrac{2\pi R^2 \Sigma}{\dot{M}}\ ,
\end{equation}
where $\Sigma$ is the disk surface density.  Taking
\response{$\Sigma \simeq 2\times 10^3~\g~\cm^{-2}$} at $R = 0.3~\au$,
one gets $\tau_{\rm wind} \sim 10^8~\yr$, which is
significantly longer than the disk lifetime
($\lesssim 10^7~\yr$), and
%dwarfs the importance of photoevaporative wind mass loss.
\response{makes photoevaporative wind mass-loss
  insignificant relatively to the total.}  The other model
is bound as the energy injection by the $h\nu = 30~\eV$
photons is insufficient for the gas to escape.

For both models, we calculate the radial distribution of
[\ion{O}{i}] $6300~\ang$ emissions via calculating the
neutral fraction of oxygen (eq.~\ref{eq:chem-bal-oxy}). Both
models have most of their [\ion{O}{i}] emission concentrated
near the center ($r\lesssim 0.5~\au$), and the unbound model
appears to be able to yield sufficient [\ion{O}{i}] emission
within the $R\lesssim 1~\au$ spatial range
($\sim 10^{-5}\;L_\odot$ for the unbound model, and
$\sim 10^{-6}\;L_\odot$ for the bound model). Nevertheless,
the mass-loss rates are tiny or even vanishing, with our
simplifications over-estimating the $\dot{M}$. 
The velocities of gas with considerable [\ion{O}{i}] 6300
emission, which are lower than the local sound speed, are
also slower than adequate to explain the observed
blueshifts and line widths of the low velocity components
\citep[e.g.][] {2018ApJ...868...28F, 2023ApJ...945..112F,
    2023NatAs...7..905F}.

One may find some subtle agreements between the
above results and Figure 3 in R23. The claimed
[\ion{O}{i}] emission in R23 comes from a sub-au region
with a negligible velocity field, with winds launched just
outside and inside this [\ion{O}{i}] emitting region. This
proposal, nevertheless, resolves neither the dilemma of
neutral oxygen versus free electrons nor the problem of
excessive cooling. Since the debate mainly focuses on the
wind launching, the proposed nearly static region is still
inadequate to conclude that the [\ion{O}{i}] emission is
consistent with photoevaporative winds if one does not
ignore the mass conservation and the sources of
streamlines.

\begin{figure*}
  \centering
  \includegraphics[width=18.2cm]
  {\figdir/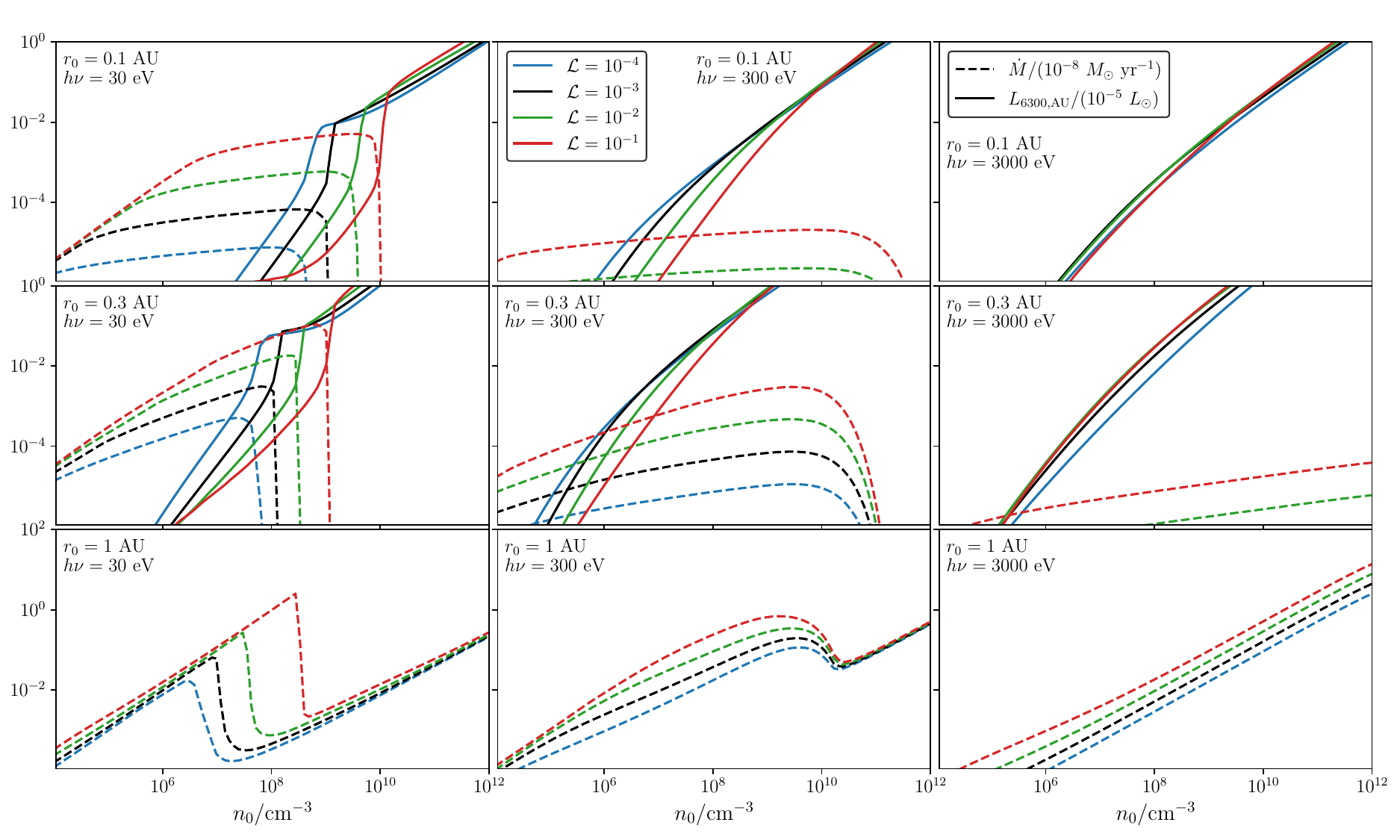}
  \caption{Photoevaporative mass-loss rates ($\dot{M}$,
    dashed lines) and 
    [\ion{O}{i}] $6300~\ang$ luminosities (solid
      lines) within $R<1~\au$ 
    ($L_{\rm 6300, AU}$) with different physical parameters:
    inner boundary radius $r_0$, inner boundary hydrogen
    nuclei number density $n_0$, XUV luminosity
    $\mathcal{L}\equiv L_{\rm XUV} / L_\odot$, and XUV
    photon energy $h\nu$. Different $\mathcal{L}$ are
    distinguished by colors in each panel as noted.
    %different line shapes show different physical
    %quantities: dashed lines for
    %$\dot{M}/(10^{-8}~M_\odot~\yr^{-1})$, and solid lines
    %for $L_{\rm 6300,AU }/(10^{-5}~L_\odot)$. 
    Various selections of inner radius $r_0$ and XUV photon
    energy $h\nu$ are presented in different panels. Note
    that $L_{6300,\au}$ vanishes for models with
    $r_0 = 1~\au$. }
 \label{fig:ana-var-model}
\end{figure*}

Other possible values of physical parameter are explored in
Figure~\ref{fig:ana-var-model}, showing the mass-loss rates
($\dot{M}$) and the [\ion{O}{i}] luminosities within the
innermost $1~\au$ ($L_{6300,\au}$) for various XUV
luminosities, XUV photon energies, and the radii and the
mass densities at the inner radial
boundaries. Figure~\ref{fig:ana-var-model} reveals that
almost all models with $n_0< 10^{10}~\cm^{-3}$ give
insufficient mass-loss rates even with excessive, unphysical
XUV luminosities (up to $10^{-1}~L_\odot$). In fact, such a
high density ($n_0\sim 10^{10}~\cm^{-3}$) at the wind base
is very unlikely. Lifting dense gas to the disk
surface requires significant heating below the disk
surfaces, which is almost impossible due to the
photoevaporative wind geometry and the significant cooling
at relatively high densities (see also
\S\ref{sec:chem-higher-emis}, and
Appendix~\ref{sec:ene-problems}). In addition, for
$h\nu \gtrsim 300~\eV$ photons, the coronal gas near $r_0$
is predominantly neutral (e.g.,
Figure~\ref{fig:ana-profiles}), and excessively high $n_0$
will cause considerable absorption at $r\gtrsim 0.3~\au$
(see also eq.~\ref{eq:ene-pen-netural-col}). Numerical
simulations have also confirmed that the wind should never
be so dense \citep[e.g.][]{2017ApJ...847...11W}. Even when
the wind is driven by the non-ideal magnetohydrodynamic
mechanisms and is relatively slow and dense, the wind-base
density is still typically $n_0 \lesssim 10^8~\cm^{-3}$
(e.g.~\citealt {2017ApJ...845...75B,
  2019ApJ...874...90W}). Therefore, the gas should never be
as dense as $n_0\gtrsim 10^{10}~\cm^{-3}$ at the beginning
of the flux tube in the disk corona or even near the disk
surface, and the mass-loss rates within $1~\au$
should be insufficient (i.e.,
$\ll 10^{-9}~M_\odot~\yr^{-1}$) in the first place.  Such
results are also consistent with the estimations in
\S\ref{sec:ene-potential}.

One may notice a dilemma emerging from the models presented
in Figure~\ref{fig:ana-var-model}:
%a photoevporative disk cannot simultaneously yield sufficient [\ion{O}{i}] emissions within $\sim 1~\au$  and adequate mass-loss rates. 
a photoevaporative disk cannot produce sufficient
[\ion{O}{i}] emissions within $\sim 1~\au$ while maintaining
adequate mass-loss rates.  The models with $r_0 = 1~\au$ and
$h\nu\gtrsim 300 \eV$ seem to resolve such a dilemma, yet
\citet{2023NatAs...7..905F} have already indicated that the
[\ion{O}{i}] emission is concentrated within the innermost
sub-au region.  Such a dilemma confirms the arguments in
\S\ref{sec:chemistry}, \S\ref{sec:energetics}, and
especially \S\ref{sec:chem-higher-emis}. Photoionization
must make its choice between the energy injection and the
neutral fraction: an unbound model with adequate mass-loss
rates always has high ionized fractions and low emissivities
from neutral oxygen, and vice versa.
% It is reminded that these models have underestimated the
% cooling processes.
It is worth noting that these models have underestimated the
cooling processes.
% Inclusion of these complexities will further bound the gas
% or make the mass-loss rates lower,
Incorporating these complexities will further bound the gas
or reduce the mass-loss rates, confirming the dilemma with
higher confidence.

\section{Summary}
\label{sec:summary}

In this work, we delve into the inner corona, viz., the gas
above the disk surfaces in the inner regions ($R< 1~\au$) of
PPDs, explicitly focusing on several major issues that could
undermine photoevaporation models when confronted with the
observed [\ion{O}{i}] $6300~\ang$ emission from the
innermost regions of PPDs.

A series of issues emerge from the thermochemistry of inner
coronae emitting in [\ion{O}{i}] $6300~\ang$. To yield the
$L_{6300}\sim 10^{-5}~L_\odot$ luminosity observed by
e.g. \citet{2023NatAs...7..905F} within a $r = 0.5~\au$
sphere, the coronal gas has to retain sufficient free
electrons and neutral oxygen atoms simultaneously at a
relatively high ($T\gtrsim 10^4~\K$) temperature. The
parameter space allowed is pretty narrow, requiring high gas
density ($n_\H\gtrsim 10^8~\cm^{-3}$) and proper
temperature [$0.5 \lesssim (T/10^4~\K) \lesssim 2$].
% requiring photoevaporative models to turn the knobs very
% carefully. Even if fine-tuning is accepted,
These conditions pose stringent constraints on
photoevaporation models, as driving and heating the gas by
the high-energy radiation alone are quite impossible due to
excessive cooling and thermal exchange (with dust
grains). Meanwhile, the cooling power under such conditions
is no less than the total X-ray luminosity of a typical
protostar, making the issue almost impossible to resolve.
In contrast, the magnetized wind model can stay inside the
proper parameter sub-space under a reasonable selection of
physical parameters \footnote{\response{Note that the
    parameters required by MHD winds for adequate
    [\ion{O}{i}] $6300~\ang$ luminosity are different from
    photoevaporative winds due to significantly lower
    ionization fractions.}}, and the accretion power
extracted from the disk accretion is sufficient to maintain
the required thermochemical conditions.

% The coronae have to be optically thin for high-energy
% photons to reach disk surfaces and launch winds at the
% innermost regions or larger disk radii, which also

The other major issue lies in the energetics of the
outflows. We specifically note that the
  blueshifted [\ion{O}{i}] $6300~\ang$ LVC is an indication
  of non-hydrostatic coronae, and the circulation motion of
  gas powered by radiative heating inside disk coronae
  should {\it not} be described as ``photoevaporative
  winds''. Since the potential well depth at the
$R < 0.3~\au$ inner disk is deeper than $\sim 15~\eV$, X-ray
photons are required to heat the gas to adequately high
temperatures to escape. However, because of their
excessively high penetration, X-ray photons can only deposit
their energy deeply below the disk surfaces; at this point,
the thermal coupling between gas and dust grains is tight,
and heating to the required temperature is
impossible. Insufficient heating makes it impossible for the
gas to leave the disk potential well, assuming it will
experience ballistic motion after instantaneous energy
injection. Even if one assumes that the heating is
continuous along the way out from the potential well,
hydrodynamic laws will yield unphysically high density at
wind bases.%  Previous works relying on the $\xi-T$ mapping
% scheme are undermined by the detachment of hydrodynamics
% from consistent thermochemistry, which results in primarily
% isothermal outflows and double-countings of the energy
% injection, thus almost always launch outflows even if the
% conditions are unphysical.

The analyses from two different aspects tend to exclude
photoevaporation models from the explanation the
inner-region [\ion{O}{i}] $6300~\ang$ profiles. A more
complex model, in which a compact [\ion{O}{i}] emitting
hydrostatic or circulating corona co-exists with
photoevaporation winds launched at larger disk radii, cannot
bridge the [\ion{O}{i}] emission to the photoevaporative
winds from the inner disk. To add to the consistency and
completeness of the analyses, we construct comprehensive
semi-analytic models with the inclusion of both hydrodynamic
and thermochemical processes. These models confirm the
analyses on the two aspects mentioned above, especially by
demonstrating the mutual exclusion between a
photoevaporative outflow with appreciable mass-loss rates
and a luminous inner-disk [\ion{O}{i}] $6300~\ang$ emission
consistent with observations.

Some previous works view wind velocities as a
criterion for distinguishing different wind launching
mechanisms. We nevertheless point out that, due to the
analyses about the thermochemical conditions of the
coronal gas, both the high-velocity components and the
low-velocity components in the TW Hya $[\ion{O}{i}]$
emission profiles are {\it not} the indications of
photoevaporative winds. In the case of the low-velocity
components observed from transitional disks with lower gas
densities ($\lesssim 10^{-3}\times$ full PPDs) and dust
abundances, the overall dilemma of photoevaporative
mass-loss versus [\ion{O}{i}] $6300~\ang$ luminosity (see
\ref{sec:chemistry}, \S\ref{sec:ana-sol}) still exists.

In conclusion, consistent co-evolution of non-equilibrium
thermochemistry and dynamics (hydrodynamics and MHD) is
necessary for physically modeling the outflows of PPDs and
plausibly explaining the observed features of disk wind
indicators.

% After calculating the conditions required by the observed
% [\ion{O}{i}] $6300~\ang$ emission, \response{we} conclude
% that X-ray photoevaporation can {\it not} be the proper
% mechanism of PPD wind launching from the innermost sub-au
% regions. X-ray alone provides insufficient heating for
% wind launching, which is further restricted by cooling
% mechanisms, especially the dust-gas thermal
% accommodation. Even if a wind is launched, the
% temperatures as well as the X-ray photoionization required
% to unbind the gas from the star gravitationally will also
% destruct neutral oxygen atoms overwhelmingly.

\bigskip

\noindent Y. Lin and L. Wang acknowledge the computing
resources provided by the Kavli Institute for Astronomy and
Astrophysics in Peking University. We thank our colleagues
for helpful discussions: Gregory Herczeg,
Xiao Hu, Haifeng Yang, Xinyu Zheng.

\bibliography{prob_phevap.bib}
\bibliographystyle{aasjournal}

\appendix
\nopagebreak

\section{Mean free path of particles and hydrodynamics}
\label{sec:mfp}

One should prove that the hydrodynamics is applicable to the
problem, and the escape of atoms as particles should not be
relevant, to justify the arguments based on hydrodynamics.  
Even if a fraction of gas particles are
sufficiently energetic to escape, the mean-free-path of
these particles can be estimated (the subscripts ``nn'',
``nc'', and ``cc'' indicate neutral-neutral,
neutral-charged, and charged-charged collisions, and
$r_{\rm n}$ is the size of the neutral particles; see
also \citealt{DraineBook}),
\begin{equation}
  \label{eq:ene-mfp}
  \begin{split}
    & \lambda_{\rm nn} \simeq 5 \times 10^{-6}~\au
      \times \left( \frac{n}{10^7~\cm^{-3}} \right)^{-1}
      \left( \dfrac{r_{\rm n}}{\ang} \right)^{-2}, \\
    & \lambda_{\rm nc} \lesssim 5 \times 10^{-7}~\au \times
      \left( \frac{n}{10^7~\cm^{-3}} \right)^{-1} , \\
    &  \lambda_{\rm cc} \simeq 3 \times 10^{-9}~\au \times
      \left( \frac{T}{10^4~\K} \right)^2
      \left(\frac{n_e}{10^7~\cm^{-3}} \right)^{-1} .
  \end{split}
\end{equation}
These lengths are tiny compared to the physical scales,
which disable the physical picture that the ``high-energy
tail'' of the Maxwell distribution could escape, and assure
the applicability of hydrodynamics.

\section{The inefficiency of X-ray heating}
\label{sec:ene-heating}

As one may infer easily by the basic estimations (e.g.,
\S\ref{sec:ene-potential}), FUV photons are not sufficiently
energetic to launch winds from the inner disk
region, and EUV photons cannot penetrate through
an [\ion{O}{i}] emitting corona to reach the disk surface in
the first place. A natural question one may raise is whether
X-ray photons are sufficiently energetic to drive
photoevaporation from the inner disk. This appendix focuses
on the issues regarding the X-ray photons
($h\nu > 0.5~\keV$), which is also the band elaborated in
e.g. \citet{2016MNRAS.460.3472E} and R23. % More comprehensive
% analyses using semi-analytic solutions for a wider range of
% parameters are presented in \S\ref{sec:ana}.

\subsection{Penetration of the X-ray energy}

%The X-ray ionization cross sections multiplied by elemental
%abundances of hydrogen and oxygen roughly read
%(see also Figure~\ref{fig:alpha-oxygen}, based on
%\citealt{Verner+etal1996})
%\begin{equation}
%  \label{eq:ene-sigma-ph-ion}
%  \begin{split}
%     & \sigma_{\chem{H}, i} \sim 1.6 \times 10^{-23}~\cm^2
%      \times \left( \frac{h \nu}{\keV} \right)^{-3} ,\\
%    & \sigma_{\chem{O}, i} X_{\chem{O}} \sim
%      3.3\times10^{-23}~\cm^2 
%      \times
%      \left(\frac{X_{\chem{O}}}{3\times 10^{-4}}\right)
%      \left( \frac{h \nu}{\keV} \right)^{-3} .
%  \end{split}
%\end{equation}
The penetration of X-ray photons in neutral materials,
measured by the column density of neutral hydrogen, is
approximately 
(for $X_\chem{O}\simeq3\times10^{-4}$),
\begin{equation}
  \label{eq:ene-pen-netural-col}
  N_{\chem{H}} \simeq (\sigma_{\chem{O}, i} X_{\chem{O}} +
  \sigma_{\chem{H}, i})^{-1} \simeq 2 \times
  10^{22}~\cm^{-2} \times \left( \frac{h \nu}{\keV} 
  \right)^3 .
\end{equation}
Considering the density profile of a Gaussian disk,
$\rho = \rho_0 \exp (- z^2 / 2 h^2)$ ($\rho_0$ is the
mid-plane density), and assume that the materials inside the
disk are all neutral. We have conducted numerical
calculations for typical PPD density profile models (in
e.g. \citealt{2017ApJ...845...75B, 2019ApJ...874...90W}),
to find that the following rough estimates yield
reasonable approximations of the density at the $\tau = 1$
surface ($f$ is the effective flaring angle of the disk
surface),
\begin{equation}
  \label{eq:ene-col-gaussian-disk}
  n_{\H,\tau = 1}\sim \dfrac{N_\H\sin f}{h}\ .
\end{equation}
We find that choosing $\sin f\sim 1/4$ gives no more than
$\sim 30~\%$ errors, which is sufficient for the
order-of-magnitude calculations in this section.

As the density of neutral hydrogen increases rapidly when
approaching the disk mid-plane, the location where the
high-energy photons get absorbed roughly satisfies
$n_{\chem{H}} h \sim N_{\chem{H}}$. Given the equilibrium
temperature of typical passive PPDs, 
\begin{equation}
  \label{eq:ene-disk-T}
  T_\eq \simeq 511~\K \left(\frac{T_1}{511~\K}\right)
  \times
  \left( \frac{R}{0.3~\au} \right)^{-1/2},
\end{equation}
and the disk scale height
($\Omega_{\K}=\sqrt{GM_*/R^3}$ is the local Keplerian angular speed;
$c_s=\sqrt{k_{\B}T_\eq/\mu}$ is the isothermal sound speed;
$\mu\simeq 2.35~m_p$ is the average molecular
mass, in which we note that the hydrogen atom in molecules
can also absorb X-ray via photoionization),
\begin{equation}
  \label{eq:ene-disk-h}
  \begin{split}
  h & = \frac{c_s}{\Omega_{\K}} \simeq 0.0074~\au
    \times \left( \frac{R}{0.3~\au} \right)^{5/4} \\
    & \ \times \left( \frac{T_1}{511~\K} \right)^{1/2}
      \left(\frac{\mu}{2.35~m_p} \right)^{-1/2}
      \left( \frac{M_*}{M_{\odot}} \right)^{-1/2} .
  \end{split}
\end{equation}
The absorption location then roughly has the number density
of hydrogen nuclei
\begin{equation}
  \label{eq:ene-absorb-dens}
  \begin{split}
    n_{\chem{H},\tau = 1}
    & \sim \frac{N_{\chem{H}}}{4h}
      \sim 10^{11}~\cm^{-3} \times
      \left( \frac{h\nu}{\keV} \right)^3
      \left( \frac{M_*}{M_{\odot}}\right)^{1/2}
    \\
    &\ \left( \frac{\mu}{2.35~m_p} \right)^{1/2}
      \left( \frac{R}{0.3~\au} \right)^{-5/4}
      \left( \frac{T_1}{511~\K} \right)^{-1/2} .
  \end{split}
\end{equation}
Such a high density clearly indicates that the absorption
location is below the disk surface. 

\subsection{Equilibrium with cooling}

Once the X-ray photons deposit their energy
%at high gas densities 
beneath the disk surface, the gas-dust thermal
accommodation will play an important role in the
thermodynamics. The accommodation rate per gas particle per
unit time is approximately ($\sigma_{\rm dust}$ is the
geometric cross section of the dust grains, and $T_\eq$
approximates the dust temperature that may be different from
the gas temperature $T$; see e.g.,
\citealt{2001ApJ...557..736G}),
\begin{equation}
  \lambda_{\rm dust} \simeq \left(
    \frac{8 k_\B T}{\pi \mu_\chem{H}} \right)^{1/2} n_{\rm dust}
  \sigma_{\rm dust} \times 2 k_\B (T - T_\eq) .
\end{equation}
The $n_{\rm dust} \sigma_{\rm dust}$ in the equation above
can be estimated by 
$n_{\rm dust} \sigma_{\rm dust} \sim (\sigma_{\rm dust} /
  \chem{H}) n_{\chem{H}, \tau = 1}$, where the ``dust cross
section per hydrogen nucleus'' parameter typically takes
$(\sigma_{\rm dust} / \chem{H}) \sim 10^{-21}~\cm^2$
(corresponding to a dust-to-gas mass ratio $\sim 10^{-2}$;
see e.g., \citealt{2019ApJ...874...90W}).  The
mass per particle here is $\mu_\chem{H}\simeq0.5~m_p$
as we are estimating the thermal accommodation for the
particles right after the ionization.

The cooling by dust grains should be compared to the effects
of X-ray photoionization heating. Because the dust
temperature $T_{\rm dust}$ is not affected by the
irradiation (Appendix~\ref{sec:Tdust-xray}), the X-ray
heating ($\gamma_{\rm X}$) and cooling
($\lambda_{\rm dust}$)
rates per atom are roughly, 
\begin{equation}
  \label{eq:ene-dust-equil}
  \begin{split}
    & \gamma_{\rm X}
    \lesssim h \nu (X_{\chem{O}} \zeta_{\chem{O}} +
      \zeta_{\chem{H}}) \sim 4 \times 10^{-19}~\erg~\s^{-1}
      \\
    & \ \times \left( \frac{L_X}{2 \times 10^{30} ~\erg~\s^{-
      1}} \right) \left( \frac{R}{0.3~\au} \right)^{-2}
      \left( \frac{h\nu}{\keV} \right)^{-3},\\ 
    & \lambda_{\rm dust}
    \sim 4 \times 10^{-19}~\erg~\s^{-1} \times
      \left(\frac{T}{T_\eq} \right)^{1/2}
      \left(\frac{T - T_\eq}{30~\K} \right)
    \\
    & \times
      \left(\frac{R}{0.3~\au} \right)^{-3/2}      
    \left(\frac{M_*}{M_\odot} \right)^{1/2}
    \left(\frac{\sigma_{\rm dust}/\H}{10^{-21}~\cm^2}
      \right) 
      \left(\frac{h\nu}{\keV} \right)^3 .
  \end{split}
\end{equation}
%Here, the photoionization rate of oxygen,
%$\zeta_{\chem{O}}$, can be obtained with
%$\zeta_{\chem{O}}\simeq L_X\sigma_{\chem{O},i}/(4\pi
%R^2h\nu)$, and one shall use $\sigma_{\chem{H}}$ for
%$\zeta_{\chem{H}}$. 
In order to stand on the safe side,
the expression of $\gamma_{\rm X}$ overestimates the X-ray
heating by assuming that the ``surplus energy''
($h \nu - I$) fully goes into the thermal energy of the
gas. A more realistic description should include secondary
ionization processes due to the collision between energetic
electrons and other neutral atoms, which will further reduce
$\gamma_{\rm X}$ drastically.

By equating $\gamma_{\rm X}$ and $\lambda_{\rm dust}$ in
eqs.~\eqref{eq:ene-dust-equil}, the terminal temperature of
the gas at photon energy $h \nu$ can be roughly estimated
as, 
\begin{equation}
  \label{eq:ene-temp-term}
  T_{\rm term}\sim T_\eq + 30~\K 
  \times \left( \frac{R}{0.3~\au} \right)^{- 1/2}
  \left( \frac{h \nu}{\keV} \right)^{-6} ,
\end{equation}
which is approximately valid when the obtained
$(T_{\rm term}-T_\eq)$ is not greater than
$T_\eq$. X-ray irradiation with $h\nu = 1~\keV$,
raising the gas temperature only by
$T - T_\eq \sim 30 ~ \K$, is far from
being sufficient to launch outflows (see the
following\S\ref{sec:ene-potential}). The dust-gas thermal
accommodation (appearing to be ``dust cooling'') will
largely offset the maximum possible heating by X-ray. We
also note that, if the cooling by atoms and molecules are
included \citep[e.g.][]{2017ApJ...847...11W}, the X-ray
heating will become even more difficult.

At different photon energy, the physical picture of
eqs.~\eqref{eq:ene-dust-equil}, \eqref{eq:ene-temp-term} is
clear: the lower the photon energy is, the shallower the
radiation can penetrate, the weaker the dust cooling is, and
the higher the temperature can reach. This is
consistent with existing literature
\citep[e.g.][]{Ercolano+2009}, that softer ionizing
photons are more efficient to produce photoevaporation.
These estimations may have caveats for ``soft X-ray''
photons with $h\nu \lesssim 0.5~\keV$. For example,
solving eq.~\eqref{eq:ene-absorb-dens} for
$h\nu = 0.3~\keV$ gives
$n_{\H,\tau = 1}\sim 3 \times 10^9~\cm^{-3}$ and
$T_{\rm term}\simeq 10^4~\K$, which is higher than the
sublimation temperature of dust grains.
% Because of the tremendous effective heat capacity due to
% the stiff thermal equilibrium between dusts and the
% stellar radiation field (Appendix~\ref{sec:Tdust-xray}),
% however, the high gas temperatures can only erode dust
% grains relatively slowly, instead of sublimate them
% rapidly. Even if 
After all dust grains removed, other cooling mechanisms
will join the thermal balance at higher
temperatures. Considering the $n_{\H,\tau=1}$ for these
$0.3~\keV$ photons, one can verify that the ionization
fraction is low\footnote{Assuming a high ionization
  fraction leads to a
  $\lambda_{\rm recomb} \sim 3\times
  10^{-15}~\erg~\s^{-1}$ recombination cooling rate per
  hydrogen nucleus at $n_\H = 3\times 10^{9}~\cm^{-3}$ and
  $T = 2\times 10^4~\K$, which prevents the runaway
  heating.}, and obtain the cooling rate per hydrogen
nucleus $\lambda\gtrsim 10^{-13}~\erg~\s^{-1}$ at
$T\simeq 2\times 10^4~\K$ by Lyman-$\alpha$ cooling alone.
At lower temperatures such as $T\sim 10^3~\K$, the energy
is removed via the fine structure transitions. For
instance, [\ion{O}{i}] $63~\micron$ can also overwhelm the
heating given $n_{\H,\tau = 1}> 10^9~\cm^{-3}$ using the
cooling rate coefficient
$\Lambda\sim 10^{-26}~\erg~\cm^3~\s^{-1}$ at
$T\simeq 10^3~\K$, \citep{1997A&AS..125..149D,
  2021ApJ...909...38D}. In fact, the consistent method
that was used in e.g. \citet{2017ApJ...847...11W} also
yields $T\sim 10^3~\K$ near the $\tau = 1$ surface at
$R = 0.3~\au$ (not presented in this paper). As we will
see in \S\ref{sec:ene-potential}, this temperature range
($T\sim 10^3-10^4~\K$) is still insufficient to launch
winds from the $R\sim 0.3~\au$ inner disk regions with
considerable efficiency. 

\subsection{The Impact of X-ray on Dust Temperatures}
\label{sec:Tdust-xray}

Is the assumption that ``$T_{\rm dust}$ is unaffected by
the X-ray irradiation'' reasonable? % More specifically, is the
% irradiation by X-ray sublimating the dust grains? 
We note that the dust grains without irradiaion are in
thermodynamic equilibria with the diffuse infrared radiation
(originating from the central star) within the disk
\citep{1997ApJ...490..368C}. 
% (the ultimate source of such diffuse radiation is the
% central protostar's optical and infrared radiation).
For dust grain sizes no smaller than the infrared
wavelengths, the Stefan-Boltzmann law constraints the
deviation from the equilibrium temperature, leading to the
following estimates of the net cooling rates that
dust temperature converges to the dust
temperature in equilibrium with the diffuse radiation field
$T_{\rm dust, \eq}\simeq T_\eq$ (the subscript ``sb'' is for
the Stefan-Boltzmann law):
\begin{equation}
  \label{eq:ene-emis-dust}
  \begin{split}
    \lambda_{\rm sb}
    & \sim 4 \pi
  a_{\rm dust}^2 \sigma_B (T_{\rm dust}^4 -
      T_{\rm dust, \eq}^4) \\
    & \sim 0.5~\erg~\s^{-1} \times
      \left( \frac{a_{\rm dust}}{1~\micron} \right)^2
    \\
    & \quad \times
      \left( \frac{T_{\rm dust, \eq}}{511~\K}\right)^4
      \left[ \left(
      \frac{T_{\rm dust, \eq}}{T_{\rm dust}} \right)^4 -1
      \right] . 
  \end{split}
\end{equation}
The maximum amount of heating that this dust
grain is ``responsible of'' can be estimated by,
\begin{equation}
  \label{eq:ene-heat-dust}
  \begin{split}
    \gamma_{\rm dust}
    & \sim \left( \frac{\pi a_{\rm dust}^2}
      {\sigma_{\rm dust} / \chem{H}} \right) \gamma_{\rm X}
      \\
    & \sim 3 \times 10^{13} \gamma_{\rm X} \ 
  \left( \frac{a_{\rm dust}}{\micron} \right)^2
  \left( \frac{\sigma_{\rm dust} / \chem{H}}{10^{-21}~\cm^2}
  \right)^{-1} . 
  \end{split}
\end{equation}
The factor $3 \times 10^{13}$ may seem a tremendous
number, but the product with $\gamma_{\rm X}$ is still
dwarfed by the $\lambda_{\rm sb}$, which can eventually
cause a temperature raise by
$(T_{\rm dust, 0} - T_{\rm dust}) \sim 3 \times
10^{-3}~\K$. If we consider smaller dust grains,
the dependence of dust emissivity on temperature will be
stiffer (roughly $\propto T^6$). In other words, due to
the stiff thermal balance between dusts and diffuse
radiation fields in the disks, $T_{\rm dust}$ is not raised
by the X-ray irradiation.

%\section{Semi-analytic Solutions of the Irradiated Disk
%Corona}
%\label{sec:ana-appdx}

\section{Issues of detaching hydrodynamics from
  thermochemistry}
\label{sec:ene-problems}

%One may find it curious that, with all the arguments presented in Section \S\ref{sec:energetics}, why in the first place were the X-ray (and FUV with lower energy photons) driven winds successfully ``launched'' in some previous works, such as R23 and \citet{2016MNRAS.460.3472E}?
\response{Some photoevaporation models, such as R23 and \citet{2016MNRAS.460.3472E}, suffer from}
%The most prominent issue emerges from 
the detachment of
thermodynamics from hydrodynamics, using a simple mapping of
gas temperature $T$ onto the ionization parameter
$\xi \equiv F_X / n_{\chem{H}}$ \citep[$F_X$ is
  the flux of X-ray photons]{2010MNRAS.401.1415O}. Recent
$\xi-T$ method \citep[e.g.][]{Picogna+2019} has attempted to
break this degeneracy by including the column density as an
additional parameter. However, there are still complications
unable to be covered by the $\xi-T$ scheme calculated on
hydrostatic grids.

Even for the hydrostatic part, the $\xi-T$ scheme
could be problematic. Penetration column density of
ionizing photons could vary by several orders of magnitude
in different energy bands (see also
Figure~\ref{fig:alpha-oxygen}), and sometimes there could
also be complicated self-shielding and cross-shielding
effects inside the PPD coronae given the multi-band
radiation conditions of the host star (see e.g.
\citealt{2017A&A...602A.105H} and references therein). The
$\xi$ value does not contain the information of photon
energy distributions, and the actual ionization rate under
the same $\xi$ could be very different. What is more, the
thermodynamics of the gas near disk surfaces can be
susceptible to multiple cooling processes. The escape
probability of cooling photons, emitted by either
molecular coolants (\chem{H_2}, \chem{OH}, \chem{H_2O};
see also \citealt{2010ApJ...722.1793O}), or atoms and ions
(e.g., \ion{C}{ii}, \chem{O}{i}), could vary by several
orders of magnitude under the same $\xi$ and lead to very
different thermodynamic conditions. Quantifying these
processes with a simple $\xi$ could lead to unrealistic
outcomes.

Inside the wind, significant heating almost stops
due to the lack of neutral materials inside the
photoionized wind.  Since the photoevaporative winds
themselves should always be optically thin to allow
high-energy photons to reach disk surfaces, column density
corrections do not resolve the problems discussed above
either. The adiabatic expansion due to outward motion will
cause significant drops in gas temperature. The
consideration from the thermodynamic aspect can
help us quantify the error that the $\xi - T$
mapping scheme introduces to the system.  When converting
thermal energy to mechanical energy, the Carnot theorem
gives the upper limit of efficiency,
\begin{equation}
  \eta < 1 -\frac{T_{\rm low}}{T_{\rm high}},
\end{equation}
where $T_{\rm low}$ and $T_{\rm high}$ are the low-end and
high-end of the mechanism. To escape from $R = 0.3~\au$, we
can take $T_{\rm high} \sim 6 \times 10^4~\K$. 
% and estimate $T_{\rm low}$ using the terminal condition of
% the gas.
When the gas expands towards the vacuum, as a disk wind
eventually does, $T_{\rm low} \ll T_{\rm high}$, and thus
$\eta$ is close to unity: the disk wind system
converts the thermal energy into mechanical energy with high
efficiencies.
% and the $\xi-T$ mapping approach actually double-counts
% the total radiation-injected energy.
The $\xi-T$ mapping approach, in contrast, maintains largely
invariant gas temperature while increasing the mechanical
energy of the system until it's roughly equivalent to the
gas thermal energy. In other words, the total
radiation-injected energy could be up to double-counted for
adiabatic expansion, one of the fundamental mechanisms that
drive the outflow.

Analyses of scalings are helpful in revealing the underlying
physics. When a wind fluid element is sufficiently far away
from the wind base, its density scales as
$n \propto r^{-2}$, where $r$ is its radial location. At
that place, the unattenuated radiation flux also scales as
$F \propto r^{-2}$. These scalings yield
$\xi \equiv F / n \propto r^0$, i.e. the ionization
parameter is insensitive to the distance.
Insensitive to the distance, the $\xi - T$
mapping technique thus yields largely constant temperature
in the winds.  According to the Parker wind theory, systems
with a polytropic index
$\gamma\equiv \mathrm{d}\ln P/\mathrm{d}\ln\rho< 3/2$ always
launch winds, let alone this roughly isothermal condition
\citep{1958ApJ...128..664P, 2007bsw..book.....M}.
% The $\xi-T$ mapping scheme ``succeeded'' in launching
% winds by effectively setting $\gamma \simeq 1$, which will
% {\it always} launch winds no matter what photon energy or
% luminosity one uses.
Consequently, the $\xi-T$ mapping scheme, by effectively
setting $\gamma \simeq 1$, leads to a scenario in which the
winds will {\it always} be launched no matter what photon
energy or luminosity one uses. This is unrealistic, as in
reality, there should be no winds when the photon energy is
too low or when the luminosity is insufficient.  Even when
the system manages to achieve constant temperature, the
physical parameters required to launch winds with
$\sim 10^{-9}~M_\odot~\yr^{-1}$ mass-loss rates are still
mostly unphysical from the $R\lesssim 0.3~\au$ part of the
disk (see also \S\ref{sec:ene-potential},
\S\ref{sec:ana}). In the most extreme case, the wind base
(with high radiation flux and high density) and somewhere in
the launched wind (with low flux and low density) might have
the same $\xi$ value, but totally different thermochemical
conditions such as reaction and cooling rates.

% In addition, the $\xi-T$ mapping cannot properly represent
% the cooling processes.  Most radiative coolings are
% two-body processes, with $\Lambda\propto n^2$, a totally
% different scaling from a simple $\xi$.  On the other hand,
% higher density results in stronger recombination and
% attenuation of the photodissociation and photoionization
% radiation fluxes, which could potentially destroy the
% coolants and limit the cooling.  The density and radiation
% flux intensity undergo drastical variation near the wind
% base, and therefore require detailed treatments to the
% cooling processes.

% END OF DOCUMENT
\end{document}